%% file: apr-data-leakage.tex
\documentclass[10pt,conference]{IEEEtran}

\input{packages}

\makeatletter
\newcommand{\linebreakand}{%
  \end{@IEEEauthorhalign}
  \hfill\mbox{}\par
  \mbox{}\hfill\begin{@IEEEauthorhalign}
}
\makeatother

\begin{document}

\title{Are Large Language Models Memorizing\\ Bug Benchmarks?}

\input{authors}

\maketitle

\begin{abstract} 
Large Language Models (LLMs) have become integral to various software engineering tasks, including code generation, bug detection, and repair. 
To evaluate model performance in these domains, numerous bug benchmarks containing real-world bugs from software projects have been developed. 
However, a growing concern within the software engineering community is that these benchmarks may not reliably reflect true LLM performance due to the risk of data leakage. 
Despite this concern, limited research has been conducted to quantify the impact of potential leakage.

In this paper, we systematically evaluate popular LLMs to assess their susceptibility to data leakage from widely used bug benchmarks. 
To identify potential leakage, we use multiple metrics, including a study of benchmark membership within commonly used training datasets, as well as analyses of negative log-likelihood and \fivegram. 
Our findings show that certain models, in particular \cgen, exhibit significant evidence of memorization in widely used benchmarks like \defj, while newer models trained on larger datasets like \llama exhibit limited signs of leakage. 
These results highlight the need for careful benchmark selection and the adoption of robust metrics to adequately assess models capabilities.
\end{abstract}

\begingroup\renewcommand\thefootnote{*}
\footnotetext{Equal contribution}
\endgroup


\section{Introduction}
\label{sec:introduction}

Large language models (LLMs) have become ubiquitous for various software engineering tasks. 
Assessing these models' abilities in context, beyond the basic evaluations typically performed upon release (e.g., on HumanEval~\cite{chen2021humaneval}), benefits from realistic benchmarks that represent real-world software development tasks.
Two significant such tasks are bug finding, through automated fault localization (FL)~\cite{flSurvey}; and bug fixing, through automated program repair~\cite{aprCACM} (APR). 
The Software Engineering community has released numerous \textit{bug benchmarks} for evaluating success on these tasks, consisting of real bugs from open-source software projects. 
Notable such datasets include, for example, \defj~\cite{defects4j} (\java) and \bugspy~\cite{bugsinpy} (\py); similarly, ML researchers recently introduced \texttt{SWEBench}~\cite{swebench}. 

\looseness-1
However, a growing concern in software engineering research is the degree to which data leakage
compromises the evaluation of true model capability~\cite{cao2024concerneddatacontaminationassessing,kapoor2022leakagereproducibilitycrisismlbased}.
\emph{Data leakage} refers to the use of information during model training that would not normally be available during prediction, leading to inflated performance metrics that misrepresent a model's true effectiveness.
The programs and bugs in many benchmarks and solutions have been publicly accessible for years, increasing the chance they were incorporated into LLM training data. 
For instance, the widely-used \defj dataset, based on popular \java projects, was first released in 2014. 

To illustrate, consider the example shown in \Cref{fig:motivation}.
Here, we prompted \cgen (6 billion parameters) to predict the next tokens based on an incomplete snippet (highlighted in yellow) from the bug-fix file for \defj bug \#39 in the \texttt{Lang} project. 
The generated text, highlighted in blue, shows that \cgen reproduced the remainder of the solution file verbatim, including specific comments (lines 31 and 35), token by token. 
This exact reproduction strongly suggests that the model has memorized the benchmark solution. 
Such behavior underscores the need for caution when using benchmarks like \defj, as they may lead to misleading conclusions about a model’s generalizability. 

Researchers in other domains, like NLP or ML, have attempted to quantify data leakage in their datasets~\cite{xu2024benchmarkingbenchmarkleakagelarge,li2023estimatingcontaminationperplexityquantifying}.  
To the best of our knowledge, this effort has not been undertaken for popular bug datasets in Software Engineering.
Detecting leakage is challenging. LLMs are often pre-trained by organizations that do not disclose their datasets, the total volume of data, nor model parameters. 
Consequently, determining whether a benchmark has been directly included or merely mirrored in the model’s training set is difficult. 
Even verifying the presence of data only establishes that the model has seen it, not necessarily that it has memorized it, particularly when it is one training sample  among billions.

Thus, we ask: \textbf{Are large language models memorizing bug benchmarks?} We systematically evaluate popular LLMs to quantify their susceptibility to data leakage on widely used bug benchmarks and raise awareness of the risks associated with using established benchmarks, which may inadvertently inflate performance by testing memorized data.

We use multiple metrics to detect potential leakage. 
First, we investigate whether benchmark data has \emph{membership} within \thestack, a widely-used used pretraining code dataset.
Following this, we apply two core metrics for leakage from prior work~\cite{xu2024benchmarkingbenchmarkleakagelarge,li2023estimatingcontaminationperplexityquantifying}: \emph{Negative Log-Likelihood (\nll)} and \fivegram. \nll provides insight into model familiarity with code snippets; \fivegram assesses the model's ability to reproduce exact sequences. 
We apply these metrics to both well-known bug benchmarks, and a new dataset of high-quality code repositories from 2024, which we mined from \github. 
This new dataset is less likely to have appeared in models' training, which allows us to compare model performance between potentially familiar data versus likely novel code. 

Our findings suggest that older models, in particular \cgen, exhibit very high \fivegram and low \nll on benchmark data, indicating a higher likelihood of memorization. 
Our evidence suggests that newer models trained on more extensive datasets, like \llama, show less 
memorization. Nonetheless, across all metrics, and models, \defj\ --- arguably the most widely-used bug benchmark
--- consistently exhibits the highest rate of potential memorization. 
These results underscore the importance of carefully selecting benchmarks to ensure reliable evaluations.

\begin{figure}[t]
\begin{lstlisting}[escapechar=\%, frame=tb, lineskip=1pt]
              * * ommited for brevity * *
%\prompt%// no search strings found, we are done
%\prompt%if (textIndex == -1) {
%\prompt%  return text;
%\prompt%}
%\prompt%int start = 0;
%\prompt%// get a good guess on the size of the result buffer so 
%\prompt%// it doesnt have to double if it goes over a bit
%\prompt%int increase = 0;
%\prompt%// count the replacement text elements that are larger 
%\prompt%// than their corresponding text being replaced
%\codegen%for (int i = 0; i < searchList.length; i++) {
%\codegen%  if (searchList[i]==null || replacementList[i]==null) { % \tikzmark{startbrace} %
%\codegen%     continue;
%\codegen%  } % \tikzmark{endbrace} %
%\codegen%  int greater = replacementList[i].length() -
%\codegen%                searchList[i].length();
%\codegen%  if (greater > 0) {
%\codegen%    increase += 3 * greater; // assume 3 matches
%\codegen%  }
%\codegen%}
%\codegen%//have upper-bound at 20%{\color{javacommentgreen} \%}% increase, then let Java take...
\end{lstlisting}

\begin{tikzpicture}[remember picture, overlay]
    \node (A) at ([xshift=7.6cm, yshift=14pt]pic cs:startbrace) {}; 
    \node (B) at ([xshift=7.57cm, yshift=-7pt]pic cs:endbrace) {};    
    
    \draw[line width=1pt,decorate,decoration={brace,amplitude=3pt,mirror}] 
        (B) -- (A) node[midway,xshift=0.3cm,yshift=0.5cm, anchor=west,rotate=-90]{\ttfamily\scriptsize{patch}};
\end{tikzpicture}

\begin{tikzpicture}
    \draw[fill=colorblind-yellow] (0,0) rectangle (0.3,0.3);
    \node[anchor=west] at (0.4,0.15) {Prompt input to \cgen.};
    \draw[fill=colorblind-blue] (0,-0.5) rectangle (0.3,-0.2);
    \node[anchor=west] at (0.4,-0.35) {Both \cgen output and the \defj solution.};

\end{tikzpicture}

\caption{
    Excerpt from \defj (\texttt{Lang:Bug 39}). 
    Given the first lines of the function until line 11, \cgen generated lines 12 to 23, matching the benchmark solution.
    \vspace{-0.8cm}
}
\label{fig:motivation}
\end{figure}

\section{Methodology}
\label{sec:methodology}

\begin{figure*}[t]
    \centering
    \includegraphics[width=0.95\linewidth]{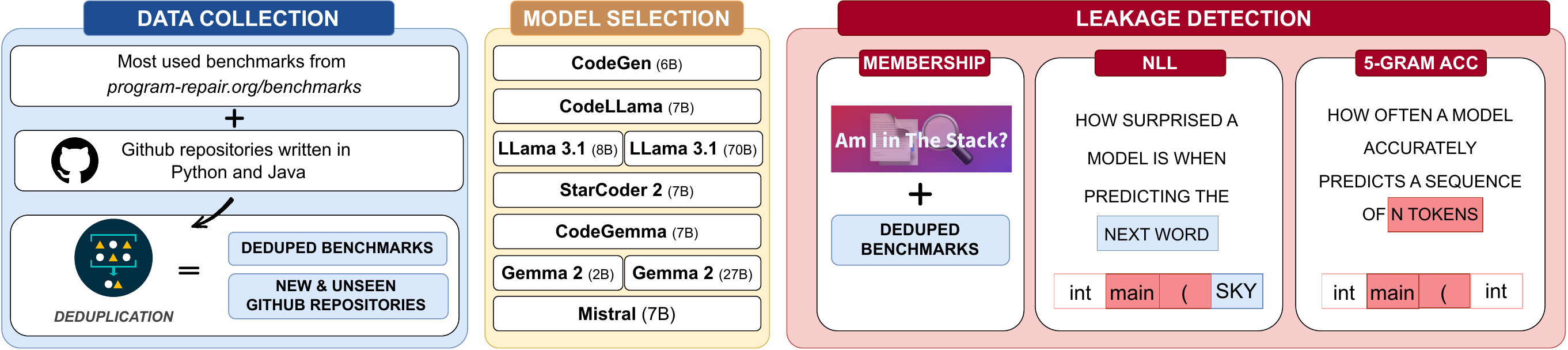}
    \caption{Overview of our methodology for detecting leakage. 
    We collected bug benchmarks and unseen repositories from 2024. 
    We evaluated \nll and $N$-gram accuracy on base models, and analyzed membership of the benchmarks in \thestack.}
    \label{fig:methodology}
\end{figure*}

\Cref{fig:methodology} overviews our methodology, which comprises
three major components: (1) data collection (\Cref{sec:data-collection}, (2) model selection (\Cref{sec:model-selection}), and (3) evaluation (\Cref{sec:evaluation}). 


\subsection{Data Collection \& Filtering.}
\label{sec:data-collection}

We select widely used bug benchmarks across common programming languages, gathering ground-truth files containing reference solutions for each bug fix. 
To provide a likely not-leaked datasets for comparison, we also curated a set of recent open-source repositories from \github. 

To collect benchmarks of interest, we reviewed \url{program-repair.org} and selected highly starred benchmarks across three programming languages:  \bugscpp~\cite{bugscpp}, \defj~\cite{defects4j}, and \bugspy~\cite{bugsinpy}.
We next included two recent datasets to serve as reference points: \gitbug~\cite{gitbug-java}, which was recently published to address the leakage issue, and \swebench~\cite{swebench}, due to its rising popularity for code-related tasks~\cite{xia2024agentlessdemystifyingllmbasedsoftware,wang2024openhandsopenplatformai,aleithan2024swebenchenhancedcodingbenchmark}. 
\Cref{tab:benchmark-details} shows details, including release year, bug count, lines of code, stars, and language. 
We choose \swebench instead of \swebenchfull due to computational constraints. 

We collected the ground truth files for each patched bug. 
To reduce computational costs and prevent bias towards particular files, we removed files with more than 85\% overlap to produce a unique sample set. 
Duplicate files appear when multiple bugs in a dataset occur in the same file (e.g., \defj in project \texttt{Lang}, Bugs 1 and 3 affect the same file). 
We kept the oldest file for consistency. 
We collected only fixed files since these correspond to the solutions a model may have memorized.

\begin{table}[tp]
    \centering
    \caption{
        Evaluation benchmark statistics, including newer and older benchmarks to assess leakage reduction over time.
    }
    \begin{tabular}{lrrrrr}
    \toprule
    \textbf{Benchmark} & \textbf{Year} & \textbf{\# Bugs} & \textbf{LOC (k)} & \textbf{\# Stars} & \textbf{Language} \\
    \midrule
    Defects4J (v1.5) & 2019 & 438 & 321 & 736 & Java \\
    BugsInPy & 2020 & 493 & 1253 & 80 & Python \\
    BugsC++ & 2021 & 209 & 4297 & 42 & C++ \\
    GitBug-Java & 2023 & 199 & 257 & 26 & Java \\
    SWEBenchLite & 2023 & 300 & 3148 & 1954 & Python \\ \midrule
    New Java Repos & 2024 & - & 132 & >100 & Java \\ 
    New Python Repos & 2024 & - & 65 & >100 & Python \\ 
    \bottomrule
    \end{tabular}
    \label{tab:benchmark-details}
\end{table}

We collected a new dataset of 3,214 \github repositories written in \java (1,183) and \py (2,031), targeting repositories from 2024 to reduce the likelihood that state-of-the-art LLMs have seen them. To focus on high-quality repositories, we applied a 100-star minimum threshold as a proxy for community interest. To ensure novelty, we used MinHash~\cite{broderMinHash} with Locality Sensitive Hashing (LSH)~\cite{gionisLSH} to filter repositories overlapping with existing training data. Due to code duplication, some files in 2024 repositories might overlap with older repositories. Therefore, we included repositories from 2022–2023 (>100 stars) to exclude 2024 repositories containing duplicated data. Finally, we randomly sampled 250 files per language to manage computational resources for evaluation.

\subsection{Model Selection.}
\label{sec:model-selection}
%




\begin{table}[]
\centering
\caption{Models used for evaluation, including their training budget in trillions of tokens, number of layers, and cutoff year.}
\label{tab:model_info}
\begin{tabular}{lrr}
\toprule
\textbf{Model} & \textbf{Tokens (T)} & \textbf{Cutoff Year} \\
\midrule
Codegen Multi (6B)~\cite{nijkamp2022codegen}    & 1  &  2022 \\
CodeLlama (7B)~\cite{rozière2024codellamaopenfoundation}        & 2.5  &  2023 \\
LlaMa 3.1 (8B / 70B)~\cite{touvron2023llama}   & 15.0 & 2024 \\
StarCoder 2 (7B)~\cite{lozhkov2024starcoder2stackv2}      & 3.5  & 2024 \\
Gemma 2 (2B / 27B)~\cite{gemmateam2024gemma2}     & 2.0 / 13.0 & 2024 \\
CodeGemma (7B)~\cite{codegemmateam2024codegemma}        & 6.5  & 2024 \\
Mistral (7B)~\cite{jiang2023mistral7b} & - & - \\
\bottomrule
\end{tabular}
\end{table}

We select a combination of models used for fault localization~\cite{yang2023largelanguagemodelstestfree}, program repair~\cite{silva2024repairllamaefficientrepresentationsfinetuned}, and vulnerability detection~\cite{zhou2024largelanguagemodelvulnerability}.
\Cref{tab:model_info} shows model information.
These models are from families of well-known code-related models, including \cgen, \llama, \gemma, \starcoder, and \mistral.

Note that we focus on open-source base models because our method requires computing the negative log-likelihood (\nll) on sequences, which is generally not possible with closed-source models. We exclude instruction-tuned models and concentrate solely on pretrained models before fine-tuning. Since instruction-tuned models are optimized for conversational formats, \ngram may be a less suitable metric for measuring memorization in these models.

\subsection{Leakage Detection}
\label{sec:leakage}

We follow strategies from prior work~\cite{xu2024benchmarkingbenchmarkleakagelarge,li2023estimatingcontaminationperplexityquantifying} to evaluate models for potential data leakage. 
\emph{Membership} operates at the repository level; \emph{Negative Log Likelihood} and \emph{N-gram accuracy}, at the file level (i.e., the compute model familiarity with a given file).  
In the bug datasets, these are the fixed (patched) files; in our novel dataset, they are randomly sampled files (Section~\ref{sec:data-collection}).

\vspace{0.5em}
\noindent\textbf{Membership}:
If a repository is included in a widely used pretraining dataset, many models have probably seen that repository's code.
We do not have direct access to, nor knowledge of, the training datasets for all evaluation models.  
However, we have partial information about the use of pretraining datasets, such as for open-source models, and closed-source models are likely to use them as well. 
We therefore assess \emph{membership} via whether a benchmark's repositories are present in 
\thestack~\cite{kocetkov2022stack3tbpermissively},
a dataset of permissibly licensed source code in 358 programming languages intended for training and fine-tuning code-based models. 
Given its size and popularity, several models report having trained on it, such as \starcoder 2~\cite{lozhkov2024starcoder2stackv2}; other closed-source models are likely to also use it. 
We used the \textit{Am I in the Stack} tool\footnote{\url{https://huggingface.co/spaces/bigcode/in-the-stack}} on each benchmark repository,
across the several versions of \thestack.

\begin{figure*}[htbp]
    \centering
    \includegraphics[width=1\textwidth]{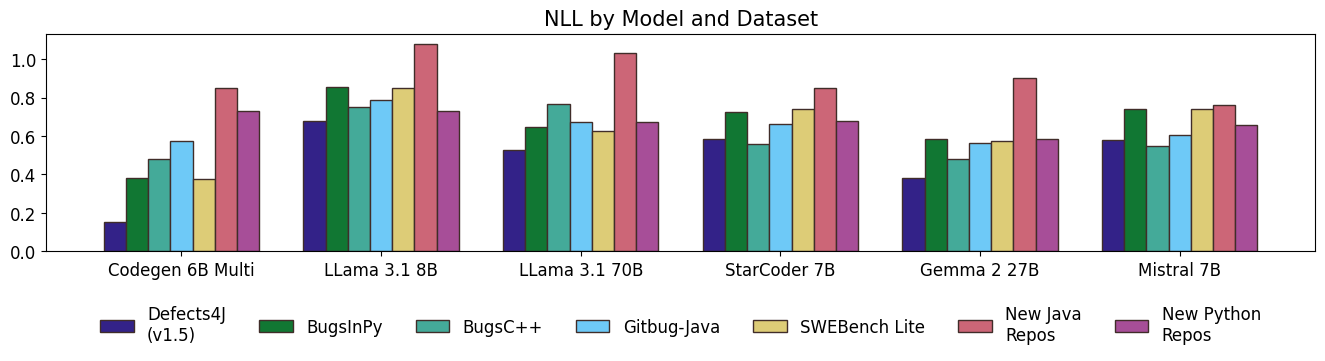} 
    \caption{\nll by model and dataset. 
    \nll is not comparable across models in different families, only across benchmarks within a family. 
    \nll for other models are consistent with the results displayed. 
    \vspace{-0.4cm}
    }
    \label{fig:nll-results}
\end{figure*}

\looseness-1
\vspace{0.5em}
\noindent\textbf{Negative Log Likelihood (\nll)}:
\nll evaluates how closely an input sequence aligns with patterns the model has learned during training in terms of 
how ``natural'' the sequence appears to the model. 
If the model has seen a data point during training, we expect it to have a lower \nll on that point compared to unseen data. 
If the model has encountered a data point \emph{many} times during training,  \nll is expected to be particularly low (i.e., close to zero) compared to arbitrary code. 

To compute \nll, we use the reference implementation publicly available on HuggingFace.\footnote{\url{https://huggingface.co/docs/transformers/en/perplexity}} 
Calculating the exact \nll for lengthy sequences is usually impractical because LLMs are trained on the limited context (moreover, we cannot fit an entire sequence in memory). 
Therefore, we split lengthy solution files into overlapping chunks, processed them individually, and combined them using a striding technique. 
We use strides of $512$ tokens when a sequence does not fit into the model's context window.





\vspace{0.5em}
\noindent\textbf{N-gram accuracy}:
$N$-gram accuracy measures the extent to which a model's output exactly matches a reference sequence at the level of $n$-grams (i.e., contiguous sequences of $n$ tokens).
High $n$-gram accuracy indicates that the model's output closely resembles the reference text, suggesting memorization. 
$N$-gram accuracy of 1.0 indicates the model can produce a sequence verbatim.
%

We follow prior work~\cite{xu2024benchmarkingbenchmarkleakagelarge} and use $5$-grams ($5$-grams strike a balance between compute efficiency and metric accuracy). 
Since most files cannot fit the context window, we use striding to cover the entire sequence. Following Xu et al.~\cite{xu2024benchmarkingbenchmarkleakagelarge}, we compute $5$-grams from five uniformly distributed starting points per stride. 
For each starting point, we provide the model with the preceding context, and check whether the predicted string matches ground truth.

\section{Results}
\label{sec:evaluation}
\vspace{-0.275em}

This section presents results assessing possible leakage of bug benchmarks in base models, using the metrics described in \Cref{sec:leakage}: membership in \thestack (Section~\ref{sec:results-membership}), Negative Log Likelihood (Section~\ref{sec:results-nll}), and \fivegram (Section~\ref{sec:nesults-ngram}). We also perform a regression analysis of model characteristics, NLL, and \fivegram (Section~\ref{sec:regression}) to better understand characteristics of models that influence data leakage.
(Note that we subsequently discuss implications in Section~\ref{sec:discussion}, and limitations and threats to the validity of our experiments in Section~\ref{sec:threats}.)



\subsection{Membership }
\label{sec:results-membership}

\begin{table}[tp]
    \centering
    \caption{Percentage of repositories in each benchmark leaked in \thestack versions 1.0, 2.0 and 2.1.}
    \begin{tabular}{lrrr}
        \toprule
        Benchmark & v1.0 (\%) & v2.0 (\%) & v2.1 (\%) \\
        \midrule
        GitBug-Java & 61.1 & 42.6 & 38.9 \\     
        BugsInPy & 94.1 & 64.7 & 64.7 \\       
        BugsC++ & 60.9 & 60.9 & 65.2 \\       
        Defects4J & 80.0 & 80.0 & 80.0 \\       
        SWEBench-Lite & 83.3 & 91.7 & 83.3 \\       

    \bottomrule
    \end{tabular}
    \label{tab:stack-membership}
\end{table}

\Cref{tab:stack-membership} shows benchmark membership in three versions of \thestack.\footnote{V1.0 is the initial 3TB of permissively licensed code, 2.0 expands to 15TB of code, and V2.1 eliminates ``opt-out" data.} 
The table excludes our new Java and Python data from 2024, as \thestack only includes data to 2023. 
Of all repositories, the new \gitbug benchmark has the lowest membership. 
\thestack contains high proportions of \defj and \swebench. 

While membership does not necessarily mean a model trained on \thestack has seen a specific fixed file (e.g., if a bug-fixing patch was applied after the dataset's cut-off date), the model may still be familiar with a project's source code. 
This familiarity could lead to higher-quality patches or results. 
This is not inherently problematic but is a critical factor to consider when assessing the model's potential for generalization.

\subsection{Negative Log Likelihood}
\label{sec:results-nll}

\begin{figure*}[hbt!]
\begin{subfigure}{.32\linewidth}
  \includegraphics[width=\linewidth]{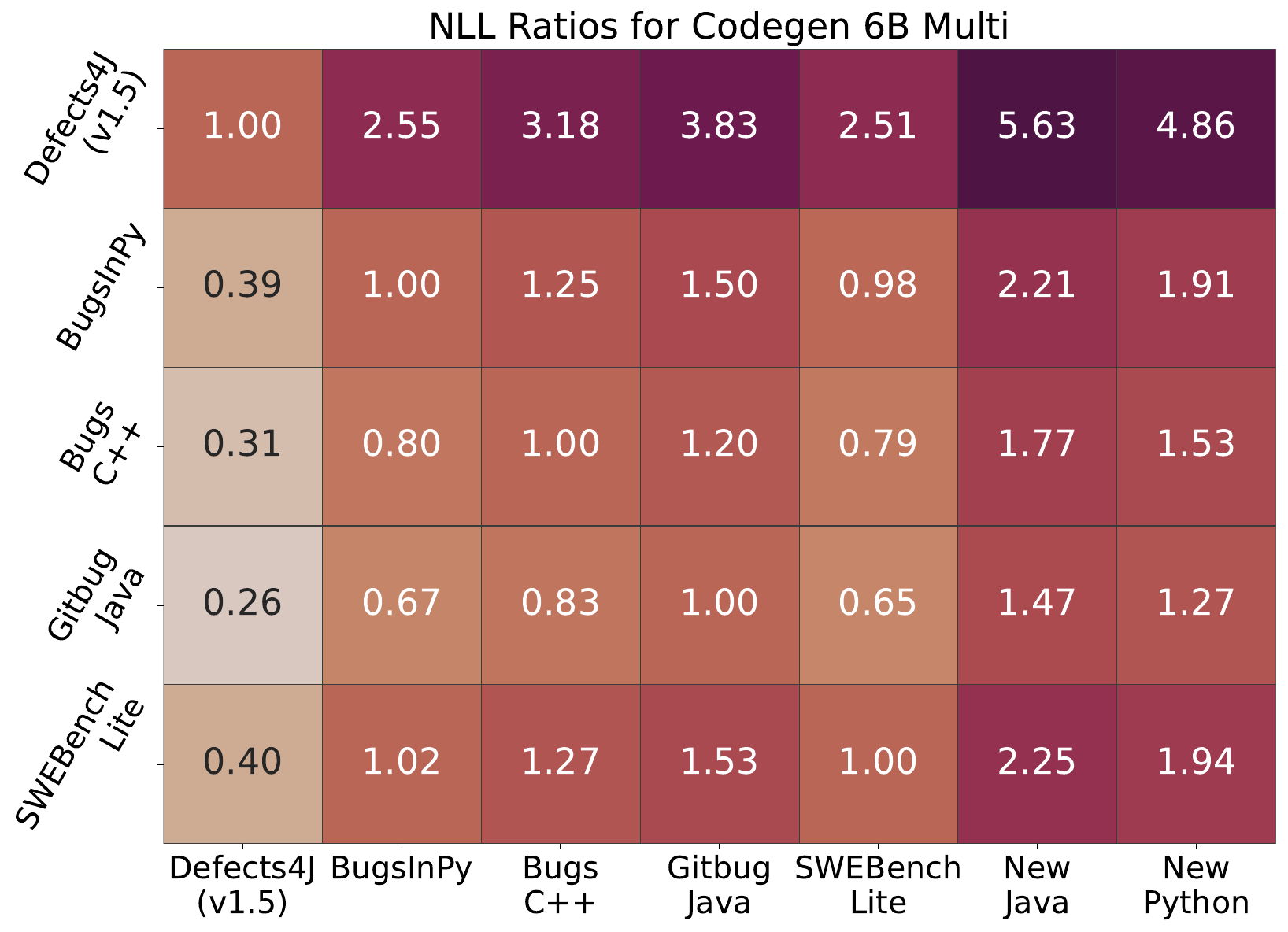}
  \label{MLEDdet}
\end{subfigure} 
\begin{subfigure}{.32\linewidth}
  \includegraphics[width=\linewidth]{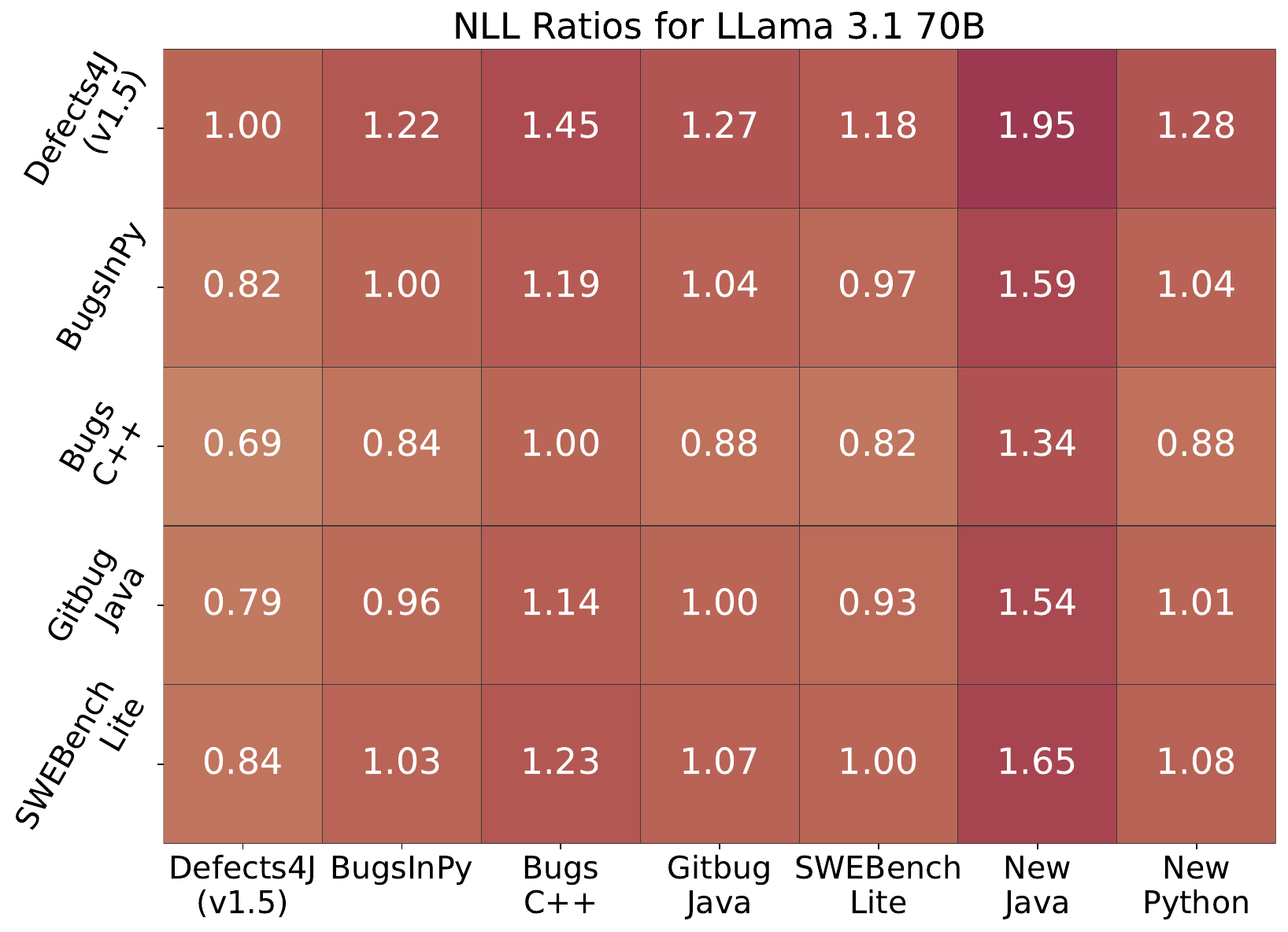}
  \label{energydetPSK}
\end{subfigure}
\begin{subfigure}{.36\linewidth}
  \includegraphics[width=\linewidth]{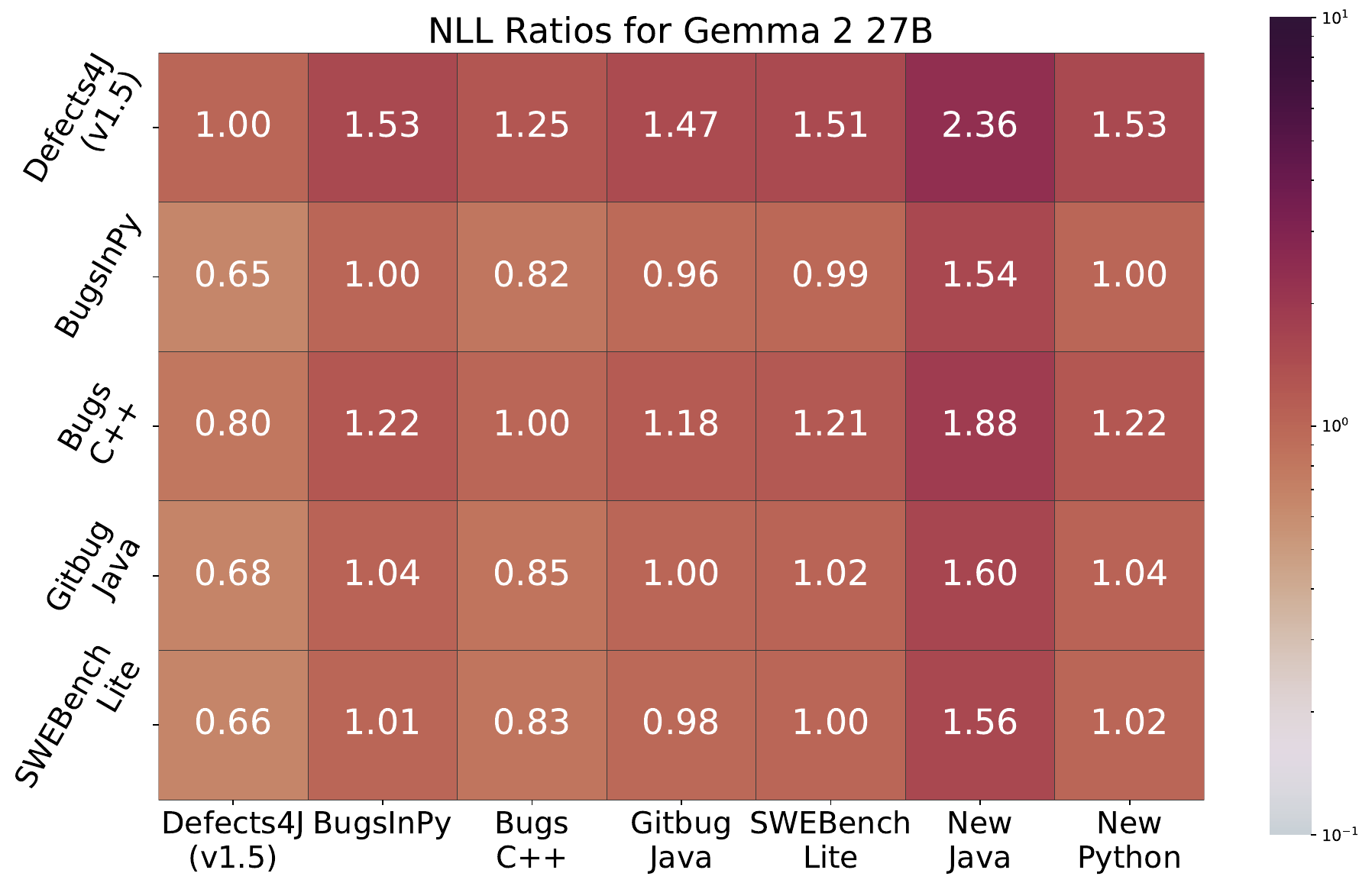}
  \label{velcomp}
\end{subfigure} 

\caption{
Heatmap illustrating the relative \nll ratios across datasets for the \cgen, \llama, and \gemma. Each cell represents the ratio of the \nll for the dataset in the column to that of the dataset in the row. 
For example, the \nll for new \java repos is $5.63\times$ higher than that for \defj. Darker colors correspond to higher ratios. \vspace{-0.4cm}
}
\label{fig:heatmap}
\end{figure*}

\Cref{fig:nll-results} shows \nll values for families of open-source models, allowing us to examine trends in familiarity across benchmarks.
Note that Negative Log Likelihood (\nll) depends on tokenization and architecture. This means we can only directly compare \nll values within model families. 

\Cref{fig:nll-results} shows that \defj consistently has the lowest \nll across all models. 
This strongly suggests 
potential data leakage. 
This is particularly evident with \cgen, which has very low \nll  (0.15) for \defj. This matches our observations in \Cref{fig:motivation} and suggests that \cgen has memorized the \defj solutions. 
We observe comparably low \nll (0.38) on the \gemma \texttt{27B} model for \defj relative to other benchmarks and repositories. 

\looseness-1
Interestingly, \cgen \texttt{6B} exhibits low \nll (0.38) on \swebench compared to other benchmarks and new data, despite being the oldest model in our evaluation, trained on older data, and the fact that \swebench was published recently.
This is because the projects in the benchmark existed prior to benchmark publication, as we also saw in the membership analysis (\Cref{tab:stack-membership}). Moreover, although \swebench is a new benchmark, the bug fixes date as early as 2017.

For all other models, the \nll values are fairly consistent across non-\defj benchmarks. 
As expected, the new repositories we collected exhibit higher \nll compared to \defj, \bugscpp, \bugspy, and \swebench. Evaluation benchmarks are derived from prominent projects and may have been seen at a pretraining time multiple times, unlike our new repositories, which likely were not. Note, however, a potential confound, which is that our new repositories may be different in distribution compared to the models' training data, which may contribute to higher \nll.  

\Cref{fig:heatmap} visualizes how each benchmark compares against other benchmarks and repositories for each model family, demonstrating how much more `familiar' a particular model is with a benchmark compared to the others.
For example, \cgen's \nll on \defj is $5.63\times$ lower than on new \java repositories, and $3.82\times$ lower than on \gitbug. 
This ratio highlights a significant level of familiarity with \defj compared to new repositories. 
Newer models, particularly \llama, exhibit relatively consistent \nll values across all benchmarks. For example, the \nll ratio between \defj and \gitbug for \llama 70B is only $1.27$, indicating that \llama perceives \defj patches as only slightly more predictable than \gitbug.
This is expected, as the newer \llama family of models was trained on significantly more data and is thus less prone to data memorization. Specifically, the \llama family was trained on $30\times$ more tokens than \cgen. 


\subsection{5-Gram Accuracy}
\label{sec:nesults-ngram}

\begin{figure*}[htbp]
    \centering
    \includegraphics[width=1\textwidth]{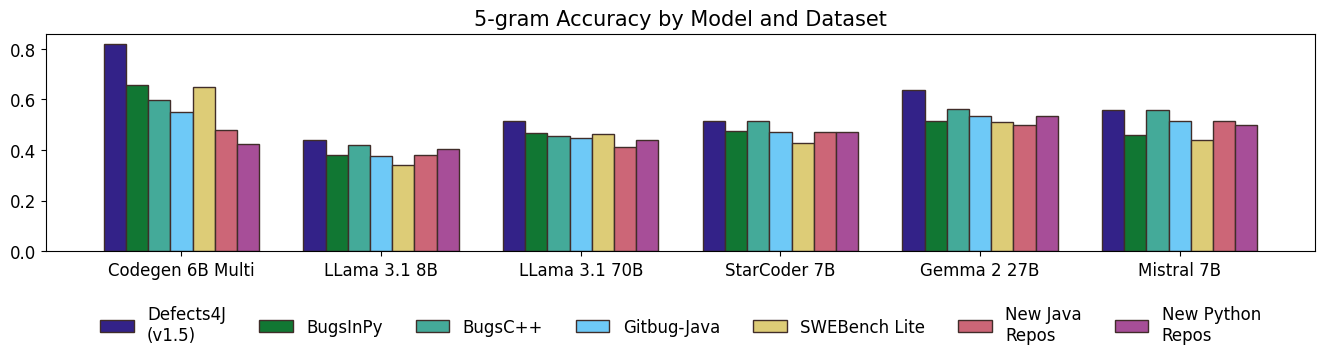} 
    \caption{\fivegram by model and dataset. Due to space constraints, we selected a sample of the most relevant models. \fivegram for other models are consistent with the results displayed.
    }
    \label{fig:ngram-results}
    \vspace{-10pt}
\end{figure*}

\Cref{fig:ngram-results} shows \fivegram results, a complementary assessment of potential model memorization (full tables, including other models are in the appendix). Note that, due to differences in model vocabularies and tokenization, interpretation of \fivegram can differ across models. 

\defj consistently exhibits the highest \fivegram across all model families. Conversely, \fivegram on \gitbug is relatively similar to that of new repositories for all models, which aligns with the expectation that most repositories in the \gitbug benchmark were not included in pretraining data (as detailed in \Cref{tab:benchmark-details}). 
For example, both \cgen and \gemma show significantly higher $5$-gram match differences between \defj and new repositories (i.e., $34$ \text{percentage points} and $14$ \text{percentage points}, respectively). 
Moreover, \cgen achieves $82\%$  \fivegram on \defj, strongly suggesting that it has likely memorized much of the benchmark's solutions. 

As expected, new repositories generally exhibit lower \fivegram across all models, with averages of 47\% and 48\% for \java and \py, respectively. These values happen likely due to the presence of common coding patterns~\cite{gabel2010Uniqueness}. Conversely, and in line with our \nll findings, \cgen shows a notably high \fivegram on \swebench, even though it is a recently published dataset, as it incorporates~older data.

When it comes to \llama family, both \llama 70B and 8B exhibit consistent \fivegram across benchmarks, which could suggest that these models are less prone to memorization due to their exposure to substantially larger datasets during pretraining. 
However, it is important to notice that \fivegram alone may not conclusively prove an absence of memorization (contrary to high $n$-gram accuracy which reliably indicates strong pattern retention). 
For example, \Cref{fig:codegenexamplellama} shows an example where \llama 70B is prompted to predict a patch from \bugspy (\texttt{fastapi}:Bug 12). 
On this file, one of the expected $5$-grams is ``\texttt{\_\_init\_\_(self,}" , and here \llama predicts ``\texttt{\_\_init\_\_(\textbackslash nself,}" instead.
This causes the $5$-grams not to match, even though the content is the same. Therefore, it is crucial to evaluate models by considering \nll, \fivegram, and membership as whole, as no single metric provides a complete picture of data leakage.

\subsection{How Do Model Characteristics Influence Risk of Leakage?}
\label{sec:regression}
%
To gain deeper insights, and using the data collected during the study, we estimate regression models for the average values of \nll and \fivegram. The regressions allow us to explore relationships that metrics alone cannot reveal. Using regression analysis, we can identify factors such as training parameters and budget as significant influences on these metrics.
%
Therefore, we use a simple mixed-effects linear model to predict the average \nll and \fivegram. As predictors, we include the models' number of \textit{parameters} and number of tokens used during pretraining (i.e., \textit{training budget}), which serves as a proxy for the unique token count. We acknowledge that reporting the exact number of unique tokens would be a more precise metric, but such data is often unavailable. We also account for potential variability caused by differences in the datasets and tokenizers by including \textit{dataset} and \textit{tokenizer type} as random effect. \Cref{tab:regression} shows results.  

In a linear regression, the intercept represents the predicted value of the dependent variable (averages of \nll and \fivegram) when all predictors are at their reference values. To make the interpretation easier, we centered the predictors around \cgen values. Therefore, the reference value of \textit{parameters} is 6B and the reference value of \textit{training budget} is 1T tokens. The regression coefficients represent the change in the dependent variable for a one-unit increase in the predictor, holding other predictors constant. 

For \nll, the predicted average is 0.744 when at reference level, i.e., parameters = 6B and training budget = 1T. For every 1B increase in number of parameters (above 6B), the predicted \nll average decreases by $0.002$ units (while holding training budget constant, at 1T). 
Similarly, for every 1T tokens increase in training budget, predicted \nll decreases by $0.014$ units. 
%
For \fivegram, the predicted average is 0.465 when predictors are at reference level. For every 1B increase in number of parameters, the predicted \fivegram increases by $0.001$ units. Similarly, a 1-unit increase in the training budget leads to an increase of $0.006$ in \fivegram. All reported coefficients have statistical significance.


Regression results reveal consistent trends across model families. For example, both \llama 8B and 70B were trained on the same data with a training budget of 15T tokens. However, the 70B model is approximately ten times larger than the 8B model, leading to an overall increase in \fivegram across all benchmarks (as shown in \Cref{fig:ngram-results}) and decrease in \nll. Similarly, within the \gemma family, the 2B model was trained on 3T tokens, while the 27B was trained on 12T. Here, we observe the same trend: average \fivegram increases in the 27B model and \nll decreases. 

%
Regression results also imply that models with more parameters tend to exhibit higher \ngram and, consequently, memorize more. For example, \Cref{fig:codegenexamplellama} shows that \llama 70B, which has the same training budget as \llama 8B, accurately predicts strings, class names, and if-statements in contexts where these predictions might not be immediately apparent, which may indicate memorization. 

\begin{table}[t]
\centering
\caption{Summary of regressions tests for negative log likelihood (\nll) and \fivegram. We report the coefficient estimates with their standard errors in parentheses. }
\begin{tabular}{lrr}
 & \multicolumn{1}{c}{\textbf{\nll}} & \multicolumn{1}{c}{\textbf{5-gram}} \\
 \toprule
\textbf{Intercept}       & 0.744 (0.094) ***                    & 0.465 (0.049)  ***\\
\midrule
\textbf{Parameters}      & -0.002 (0.001) *\phantom{**}  & $0.001$  ($4.0 e{-4}$)  *\phantom{**} \\
\textbf{Training budget} & -0.014 (0.004) **\phantom{*}        & $0.006$ ($2.4 e{-3}$) *\phantom{**} 
\\ 
\bottomrule
\\
\multicolumn{3}{l}{\textit{Note:  *** p-value \textless~0.001, ** p-value \textless~0.01, * p-value \textless~0.05}} 
\end{tabular}
\label{tab:regression}
\vspace{-1.2em}
\end{table}


\section{Discussion}

\input{figures/qualitative/example_paper}

\label{sec:discussion}


Our evaluation provides compelling evidence that data leakage is an especially significant issue for \defj (V1.5). 
This is evident from the lower \nll values and higher \fivegram. In particular, \cgen achieves $82\%$ \fivegram on \defj, while both \texttt{CodeLlama 7B} and \texttt{Gemma-2 27B} attain $64\%$.  Moreover, given that \defj is incorporated into the widely-used pretraining dataset (\thestack), with $80\%$ membership, eliminating this leakage in future models is likely nearly impossible. We also observe similarly low \nll values in \swebench for \cgen.

Newer benchmarks \bugspy and \bugscpp exhibit lower leakage risk in almost all models. A smaller percentage of their repositories are indexed in the latest version of \thestack. While \bugspy and \bugscpp exhibit slightly lower \nll compared to new repositories, their \fivegram and \nll values are not significantly different from newer, more-likely-unseen benchmarks like \gitbug. 
This suggests that, for now, researchers can use these benchmarks with a relatively low risk of data leakage.
However, as pretraining datasets continue to evolve, we recommend that researchers regularly assess \thestack membership, \fivegram and \nll values of these benchmarks, especially compared to more recent data, to monitor and mitigate potential data contamination. 

We also observe that the \llama family seems to exhibit lower memorization of benchmark solutions. Nonetheless, we still observed cases where \llama 70B outputs solution files despite little context (e.g., \Cref{fig:example_ngrams_semantic_memorization}).


We suggest researchers consider supplementing their evaluations with more recent benchmarks such as \gitbug.
Benchmarks like \gitbug, which focus on recent bugs and patches, are less likely to have been included in pretraining datasets compared to established benchmarks. 
Leveraging these newer benchmarks can provide more reliable evaluations for assessing model's capabilities.

\section{Limitations and Threats}
\label{sec:threats}

\noindent\textbf{Data Collection:} Despite efforts to filter out older \github repositories when collecting new data, we anecdotally observed instances where files appeared to be adaptations of existing files (e.g., from 2018) Our filtering process may not have perfectly excluded legacy code. 
We moreover cannot guarantee that the new repositories are identically distributed compared to those in the benchmarks. 
To mitigate this issue, given that the repositories in the benchmarks tend to be highly recognizable projects, we applied a >100-star filter to argue that the selected projects are of comparable quality.

\vspace{0.5em}
\noindent\textbf{Train + Test Splits}: 
For benchmarks like \defj (V1.5) and \bugspy, the patch files and buggy files are very similar (typically a patch involves only changing a small number of lines). 
LLMs may have only seen the train split 
of these benchmarks at pretraining time. 
This would result in high \fivegram and low \nll on patch files, even if only buggy files were leaked. 
Nonetheless, we observed multiple cases where models output patched files verbatim. We mitigate this by looking at trends in NLL and \fivegram rather than absolute numbers in our analysis.

\vspace{0.5em}
\noindent\textbf{Forgetting}: The findings presented in this paper primarily address leakage in the context of base models. Empirical results show that models can ``forget'' portions of their pretraining data during fine-tuning~\cite{luo2024empiricalstudycatastrophicforgetting}. 
That said, while full-scale model fine-tuning may reduce leakage risks, more recent fine-tuning strategies---such as the addition of adapter layers---often ``freeze'' pretrained weights. 
This practice can inadvertently increase the likelihood of data leakage. 
Furthermore, larger models exhibit a stronger tendency to memorize training data~\cite{tirumala2022memorizationoverfittinganalyzingtraining}, suggesting that while fine-tuning may mitigate leakage for smaller models, it is more likely to exacerbate leakage concerns in larger models. 
We leave empirical assessment of these risks to future work.


\section{Related Work}

\noindent\textbf{Large Language Models (LLM):}
LLMs have shown promise across a wide range of natural language~\cite{rajpurkar2016squad100000questionsmachine,hendrycks2021measuringmassivemultitasklanguage} and code generation tasks~\cite{chen2021evaluatinglargelanguagemodels, austin2021programsynthesislargelanguage, hendrycks2021measuringcodingchallengecompetence}. Current state of the art LLMs include open source models such as \cgen, \codellama, \llama, \gemma and  and closed-source models such as \texttt{GPT-4o}, \texttt{Claude 3.5}, and \texttt{Gemini 1.5}. LLMs have shown state of the art performance in APR, suggesting patches to buggy code segments, detecting vulnerabilities and helping pinpoint buggy lines in a piece of code~\cite{yang2023largelanguagemodelstestfree, silva2024repairllamaefficientrepresentationsfinetuned, xia2023conversationgoingfixing162}.

Despite the widespread success of LLMs in APR tasks, there are still significant concerns regarding data leakage. This is prevalent for LLMs trained on large, publicly available code repositories, as these models might "remember" solutions from benchmark datasets, resulting in inflated performance. 

\vspace{0.5em}
\noindent\textbf{Measuring Data Leakage:} This task is still an open challenge. Three common ways to measure data leakage are perplexity or \nll, \ngram, and model performance. \nll based approaches ~\cite{li2023estimatingcontaminationperplexityquantifying, xu2024benchmarkingbenchmarkleakagelarge} quantify leakage by comparing loss across benchmarks. \ngram approaches~\cite{xu2024benchmarkingbenchmarkleakagelarge,shi2024detectingpretrainingdatalarge}, prompt a model with context and measure the overlap of generated code with the ground truth. Finally, performance approaches~\cite{jain2024livecodebenchholisticcontaminationfree, zuoLeakage} measure the difference in performance between a benchmark and a transformed version of the benchmark. While we build on established techniques from the literature, our work is the first to apply these metrics to bug benchmarks widely recognized in the software engineering community. Additionally, we conduct a comprehensive study of both recent models, such as the \llama family, and earlier models like \cgen.

\section{Conclusion}

In this paper, we measure data leakage risks across multiple widely used bug benchmarks and state-of-the-art open-source models. Older benchmarks, especially \defj, exhibit higher memorization signals across all models, particularly in \cgen 6B. Newer datasets such as \gitbug, \bugspy, \bugscpp show lower leakage risk, with similar \nll and \fivegram to new 2024 repositories. We also find that newer models with higher training budgets display significantly lower risks of data leakage. Our findings suggest that researchers should consider both the models they use in evaluation and pair older benchmarks, such as \defj, with more recent benchmarks, such as \gitbug, to improve model evaluation and address potential leakage issues 

\section{Acknowledgments}
The authors would like to thank Luke Dramko for providing code that we used for deduplicating 2024 repositories. 
This work was supported by Fundação para a Ci\^encia e Tecnologia (FCT) through the CMU Portugal Dual PhD program: Daniel Ramos (SFRH/BD/150688/2020), Claudia Mamede (PRT/BD/155043/2022), Paulo Canelas (SFRH/BD/151469/2021), Catarina Gamboa (PRT/BD/154254/2021).

\bibliographystyle{IEEETran}
\bibliography{references}

\input{appendix}

\end{document}

%% file: packages.tex
\usepackage{amsmath,amsfonts}
\usepackage{algorithmic}

\usepackage{csquotes}
\usepackage{graphicx}
\usepackage{booktabs} 
\usepackage{colortbl}
\usepackage[dvipsnames,table]{xcolor}
\usepackage{multirow}
\usepackage{subcaption}  

\usepackage{tikz}
\usetikzlibrary{tikzmark}
\usetikzlibrary{decorations.pathreplacing} 

\usepackage{hyperref}
\usepackage{cleveref}

\usepackage{float}
\usepackage{xspace}

\usepackage{etoolbox} 
\let\oldttfamily\ttfamily 
\renewcommand{\ttfamily}{\oldttfamily\footnotesize}

\usepackage[normalem]{ulem}

\usepackage{color}
\usepackage{tcolorbox}
\usepackage{titlesec}

\definecolor{antiquefuchsia}{rgb}{0.57, 0.36, 0.51}
\definecolor{asparagus}{rgb}{0.53, 0.66, 0.42}
\definecolor{githubgreen}{RGB}{200, 255, 200}  
\definecolor{keywordpurple}{HTML}{700080}

\definecolor{colorblind-blue}{HTML}{CFE1F3}
\definecolor{colorblind-red}{HTML}{F4A582}
\definecolor{colorblind-dkblue}{HTML}{99CCFF}  
\definecolor{colorblind-yellow}{HTML}{FFFBD6}

\definecolor{keywordblue}{rgb}{0.0, 0.3, 0.6}    
\definecolor{stringpurple}{rgb}{0.58, 0.0, 0.82} 
\definecolor{commentgreen}{rgb}{0.0, 0.5, 0.0}   
\definecolor{builtinorange}{rgb}{0.8, 0.4, 0.0}  

\definecolor{javakeywordblue}{rgb}{0.0, 0.3, 0.7}      
\definecolor{javakeywordteal}{rgb}{0.0, 0.5, 0.4}

\definecolor{javastringpurple}{rgb}{0.58, 0.0, 0.82}   
\definecolor{javacommentgreen}{rgb}{0.0, 0.5, 0.0}     
\definecolor{javabuiltinorange}{rgb}{0.8, 0.4, 0.0}    
\definecolor{amaranth}{rgb}{0.9, 0.17, 0.31}

\usepackage{listings}
\crefname{lstlisting}{listing}{listings}
\Crefname{lstlisting}{Listing}{Listings}

\newcommand{\codegen}{\makebox[0pt][l]{\color{colorblind-blue}\rule[-4pt]{\linewidth}{10pt}}}

\newcommand{\prompt}{\makebox[0pt][l]{\color{colorblind-yellow}\rule[-4pt]{\linewidth}{10pt}}}

\definecolor{bisque}{rgb}{1.0, 0.89, 0.77}
\definecolor{burntorange}{rgb}{0.8, 0.33, 0.0}
\definecolor{cerulean}{rgb}{0.0, 0.48, 0.65}
\definecolor{cordovan}{rgb}{0.54, 0.25, 0.27}
\definecolor{lightergray}{rgb}{0.95,0.95,0.95}

\newcommand{\generated}[1]{\colorbox{colorblind-blue}{\strut\textcolor{cerulean}{#1}}}
\newcommand{\notgenerated}[1]{\colorbox{bisque}{\strut\textcolor{burntorange}{#1}}}
\usepackage{DejaVuSansMono}
\usepackage{comment}

\usepackage[T1]{fontenc}

\lstdefinestyle{javalang}{
    language=Java,
    basicstyle=\ttfamily\scriptsize,
    keywordstyle=\color{keywordpurple},
    commentstyle=\itshape\color{javacommentgreen},
    stringstyle=\color{javastringpurple},
    identifierstyle=\color{black},
    numberstyle=\scriptsize\color{gray},              
    numbers=left,                                    
    numbersep=6pt,                                    
    xleftmargin=12pt,                                
    frame=single,
    framerule=0.5pt,
    framesep=5pt,
    showstringspaces=false,
    breaklines=true,
    tabsize=2,
    captionpos=b,
    float=t,
    floatplacement=tbp,
    morekeywords={abstract, assert, boolean, break, byte, case, catch, char, class, const, continue, default, do, double, else, enum, extends, final, finally, float, for, goto, if, implements, import, instanceof, int, interface, long, native, new, null, package, private, protected, public, return, short, static, strictfp, super, switch, synchronized, this, throw, throws, transient, try, void, volatile, while},
    emph={System, out, println, String, Integer, Double, ArrayList, HashMap, HashSet},
    emphstyle=\color{javabuiltinorange},
    escapeinside={(*@}{@*)},  
    moredelim=**[is][\colorbox{githubgreen}]{@+}{@+}  
}
\lstdefinestyle{pythonlang}{
    language=Python,
    basicstyle=\ttfamily\scriptsize,
    keywordstyle=\color{keywordpurple},
    commentstyle=\itshape\color{javacommentgreen},
    stringstyle=\color{javastringpurple},
    identifierstyle=\color{black},
    numberstyle=\scriptsize\color{gray},              
    numbers=left,                                    
    numbersep=6pt,                                    
    xleftmargin=12pt,                                
    frame=single,
    framerule=0.5pt,
    framesep=5pt,
    showstringspaces=false,
    breaklines=true,
    tabsize=4,                                       
    captionpos=b,
    float=t,
    floatplacement=tbp,
    morekeywords={and, as, assert, async, await, break, class, continue, def, del, elif, else, except, False, finally, for, from, global, if, import, in, is, lambda, None, nonlocal, not, or, pass, raise, return, True, try, while, with, yield},
    emph={print, str, int, float, list, dict, set, tuple, open, range, len, sum, zip, map, filter, any, all, max, min},
    emphstyle=\color{javabuiltinorange},
    escapeinside={(*@}{@*)},  
    moredelim=**[is][\colorbox{githubgreen}]{@+}{@+}  
}

\lstdefinestyle{diff}{
    language=diff,
    basicstyle=\ttfamily\small,
    morecomment=[f][\color{diffstart}]{@@},
    morecomment=[f][\color{diffincl}]{+\ },
    morecomment=[f][\color{diffrem}]{-\ },
  }

\newcommand{\defj}{\texttt{Defects4J}\xspace}
\newcommand{\bugspy}{\texttt{BugsInPy}\xspace}
\newcommand{\bugscpp}{\texttt{BugsCpp}\xspace}

\newcommand{\swebench}{\texttt{SWEBench-Lite}\xspace}
\newcommand{\swebenchfull}{\texttt{SWEBench}\xspace}
\newcommand{\gitbug}{\texttt{GitBug-Java}\xspace}
\newcommand{\java}{\texttt{Java}\xspace}
\newcommand{\py}{\texttt{Python}\xspace}

\newcommand{\cgen}{\texttt{codegen-multi}\xspace}
\newcommand{\llama}{\texttt{LLaMa 3.1}\xspace}
\newcommand{\gemma}{\texttt{Gemma 2}\xspace}

\newcommand{\codellama}{\texttt{CodeLLaMa}\xspace}
\newcommand{\mistral}{\texttt{Mistral}\xspace}

\newcommand{\starcoder}{\texttt{StarCoder}\xspace}
\newcommand{\github}{\texttt{GitHub}\xspace}
\newcommand{\fivegram}{\texttt{5-gram accuracy}\xspace}
\newcommand{\ngram}{\texttt{n-gram accuracy}\xspace}

\newcommand{\thestack}{\texttt{TheStack}\xspace}
\newcommand{\nll}{\texttt{NLL}\xspace}

%% file: authors.tex
\author{
    \IEEEauthorblockN{Daniel Ramos}
    \IEEEauthorblockA{
        \textit{Carnegie Mellon University, INESC-ID}\\
        Pittsburgh, PA, USA \\
        danielrr@cmu.edu
    }
\and
    \IEEEauthorblockN{Claudia Mamede\textsuperscript{*}}
    \IEEEauthorblockA{
        \textit{Carnegie Mellon University, FEUP}\\
        Pittsburgh, PA, USA \\
        cmamede@andrew.cmu.edu
    }
\and
    \IEEEauthorblockN{Kush Jain\textsuperscript{*}}
    \IEEEauthorblockA{
        \textit{Carnegie Mellon University}\\
        Pittsburgh, PA, USA \\
        kdjain@andrew.cmu.edu
    }
\linebreakand
    \IEEEauthorblockN{Paulo Canelas\textsuperscript{*}}
    \IEEEauthorblockA{
        \textit{Carnegie Mellon University, LASIGE}\\
        Pittsburgh, PA, USA \\
        pasantos@andrew.cmu.edu
    }
\and
    \IEEEauthorblockN{Catarina Gamboa\textsuperscript{*}}
    \IEEEauthorblockA{
        \textit{Carnegie Mellon University, LASIGE}\\
        Pittsburgh, PA, USA \\
        cgamboa@andrew.cmu.edu
    }
\and
    \IEEEauthorblockN{Claire {Le Goues}}
    \IEEEauthorblockA{
        \textit{Carnegie Mellon University}\\
        Pittsburgh, PA, USA \\
        clegoues@andrew.cmu.edu
    }
}

%% file: figures/qualitative/example_paper.tex
{
\lstset{style=pythonlang}
\begin{figure*}[htbp]
    \centering
    \begin{subfigure}[b]{0.48\textwidth}
        \begin{lstlisting}[escapechar=\%, frame=tb, lineskip=1pt]
%\prompt%                * * omitted for brevity * *
%\prompt%    if scheme.lower() != "bearer":
       if self.auto_error:
          raise HTTPException(
            status_code=HTTP_403_FORBIDDEN, 
            detail="Invalid authentication 
                    %\notgenerated{credentials}%%\generated{scheme}%")
       else:
          return None
   return %\notgenerated{HTTPAuthorizationCredentials(scheme=scheme,}% 
%\notgenerated{credentials=credentials)}%%\generated{self.model.parse(credentials)}%

class HTTP%\notgenerated{Digest(HTTPBase)}%%\generated{BearerModel(BaseModel):}%:
    def __init__(self, *, %\generated{bearerFormat: str = None},%
%\codegen%                * * omitted for brevity * *
        \end{lstlisting} 
        \begin{tikzpicture}
            \draw[fill=colorblind-yellow] (0,0) rectangle (0.3,0.3);
            \node[anchor=west] at (0.4,0.12) {Prompt input to \llama \texttt{8B.}};
            \draw[fill=colorblind-blue] (0,-1.0) rectangle (0.3,-0.7);
            \node[anchor=west] at (0.4,-0.88) {Extra code generated.};
            \draw[fill=bisque] (4,-1.0) rectangle (4.3,-0.7);
            \node[anchor=west] at (4.4,-0.88) {Ground truth not generated.};
            \draw[fill=white] (0,-0.5) rectangle (0.3,-0.2);
            \node[anchor=west] at (0.4,-0.38) {Matching Ground Truth \& Predicted by \llama 8B.};
        \end{tikzpicture}
        \caption{Example of generation by \llama \texttt{8B}.}
        \label{fig:codegenexamplellama2}
    \end{subfigure}
    \hfill
    %
    %
    %
    %
    %
    %
    %
    \begin{subfigure}[b]{0.48\textwidth}
        \begin{lstlisting}[escapechar=\%, frame=tb, lineskip=1pt]
%\prompt%                * * omitted for brevity * *
%\prompt%   if scheme.lower() != "bearer":  
      if self.auto_error:
         raise HTTPException(
         status_code=HTTP_403_FORBIDDEN,%\generated{ \textbackslash n}%
         detail="Invalid authentication credentials"%\generated{\textbackslash n}%
      )
      else:
         return None
   return HTTPAuthorizationCredentials(scheme=scheme,
                             credentials=credentials)

class HTTPDigest(HTTPBase):
    def __init__(%\generated{\textbackslash n}%self,%\generated{\textbackslash n}%*,%\generated{\textbackslash n}%%\generated{qop: str = None \textbackslash n},%
%\codegen%                * * omitted for brevity * *
\end{lstlisting}
        \begin{tikzpicture}
            \draw[fill=colorblind-yellow] (0,0) rectangle (0.3,0.3);
            \node[anchor=west] at (0.4,0.12) {Prompt input to \texttt{\llama 70B}.};
            \draw[fill=colorblind-blue] (0,-1.0) rectangle (0.3,-0.7);
            \node[anchor=west] at (0.4,-0.88) {Extra code generated.};
            \draw[fill=bisque] (4,-1.0) rectangle (4.3,-0.7);
            \node[anchor=west] at (4.4,-0.88) {Ground truth not generated.};
            \draw[fill=white] (0,-0.5) rectangle (0.3,-0.2);
            \node[anchor=west] at (0.4,-0.38) {Matching Ground Truth \& Predicted by \llama 70B.};
        \end{tikzpicture}
        \caption{Example of generation by \llama 70B.}
        \label{fig:codegenexamplellama}

    \end{subfigure}
    \caption{
    Example patch of a bug from BugsInPy (\texttt{fastapi}:Bug 12). We prompt each model with the 30 lines prior to the patch. We highlight the \generated{extra code generated} not in the ground truth and \notgenerated{ground truth code not generated} by the model. The remaining lines of the examples are omitted for brevity.
    \vspace{-0.4cm}
    }
    \label{fig:example_ngrams_semantic_memorization}
    \begin{tikzpicture}[remember picture, overlay]
        \node (A) at ([xshift=16.8cm, yshift=111pt]pic cs:startbrace) {}; 
        \node (B) at ([xshift=16.8cm, yshift=55pt]pic cs:endbrace) {};    
        
        \draw[line width=1pt,decorate,decoration={brace,amplitude=3pt,mirror}] 
            (B) -- (A) node[midway,xshift=0.25cm,yshift=0.5cm, anchor=west,rotate=-90]{\ttfamily\scriptsize{patch}};
    \end{tikzpicture}
    \begin{tikzpicture}[remember picture, overlay]
        \node (A) at ([xshift=7.45cm, yshift=115pt]pic cs:startbrace) {}; 
        \node (B) at ([xshift=7.45cm, yshift=60pt]pic cs:endbrace) {};    
        
        \draw[line width=1pt,decorate,decoration={brace,amplitude=3pt,mirror}] 
            (B) -- (A) node[midway,xshift=0.25cm,yshift=0.5cm, anchor=west,rotate=-90]{\ttfamily\scriptsize{patch}};
    \end{tikzpicture}
    \vspace{-1em}
\end{figure*}
}

%% file: appendix.tex
\onecolumn    

\newpage\clearpage
\appendices

\section{}


\begin{figure}[h]
    \centering
    \includegraphics[width=0.9\textwidth]{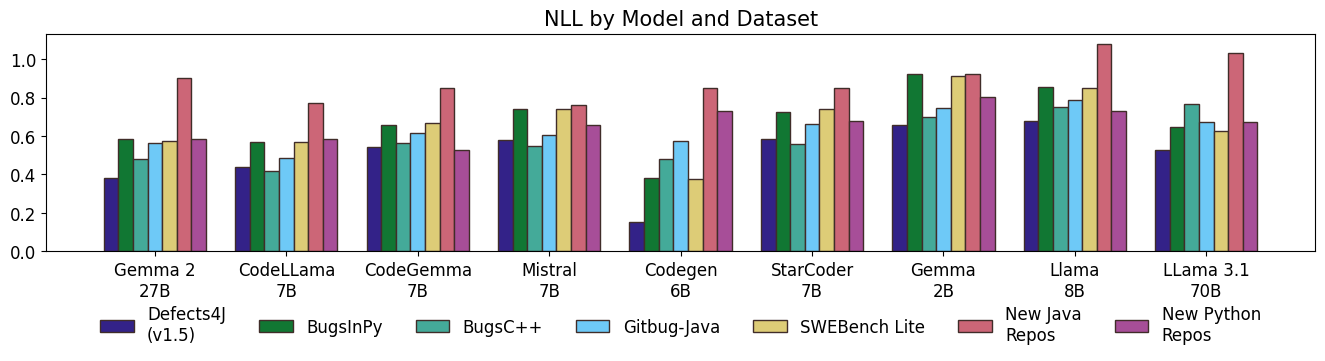}
    \caption{NLL results for all the models and benchmarks.}
\end{figure}

\begin{table*}[h]
\centering
\caption{Average \nll across benchmarks and models}
\resizebox{0.9\textwidth}{!}{%
\begin{tabular}{lrrrrrrr}
\toprule
 & Defects4J (V1.5) & BugsInPy & New Java Repos & GitBug Java & BugsCpp & SWE Bench Lite & New Python Repos \\
\midrule
Salesforce/codegen-6B-multi & 0.15 & 0.38 & 0.85 & 0.58 & 0.48 & 0.38 & 0.73 \\
meta-llama/CodeLlama-7b-hf & 0.44 & 0.57 & 0.77 & 0.48 & 0.42 & 0.57 & 0.58 \\
LlaMa-3.1-8B & 0.68 & 0.85 & 1.08 & 0.79 & 0.75 & 0.85 & 0.73 \\
starcoder2-7b & 0.58 & 0.72 & 0.85 & 0.66 & 0.56 & 0.74 & 0.68 \\
codegemma-7b & 0.54 & 0.66 & 0.85 & 0.62 & 0.57 & 0.67 & 0.53 \\
gemma-2-2b & 0.66 & 0.92 & 0.92 & 0.74 & 0.70 & 0.91 & 0.80 \\
Mistral-7B-v0.3 & 0.58 & 0.74 & 0.76 & 0.61 & 0.55 & 0.74 & 0.66 \\
Gemma 2 27b & 0.38 & 0.58 & 0.90 & 0.56 & 0.48 & 0.58 & 0.59 \\
LlaMa 70B & 0.53 & 0.65 & 1.03 & 0.67 & 0.77 & 0.63 & 0.68 \\
\bottomrule
\end{tabular}}
\end{table*}

\begin{figure*}[h]
    \centering
    \includegraphics[width=0.9\textwidth]{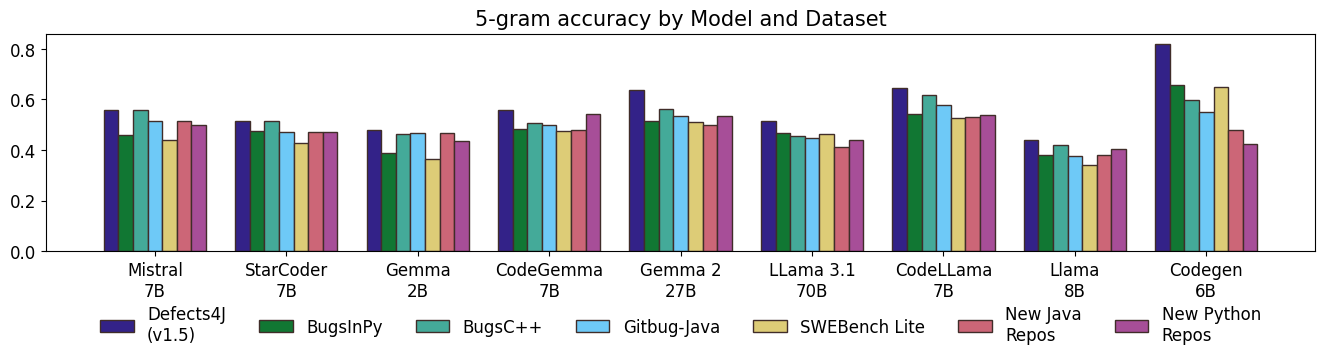}
    \caption{5-gram results for all the models and benchmarks.}
\end{figure*}


\begin{table*}[h]
\centering
\caption{Average 5-gram accuracy across multiple benchmarks}
\resizebox{0.9\textwidth}{!}{%
\begin{tabular}{lrrrrrrr}
\toprule
 & Defects4J (V1.5) & BugsInPy & New Java Repos & GitBug Java & BugsCpp & SWEBench Lite & New Python Repos \\
\midrule
Salesforce/codegen-6B-multi & 0.82 & 0.66 & 0.48 & 0.55 & 0.60 & 0.65 & 0.42 \\
meta-llama/CodeLlama-7b-hf & 0.64 & 0.54 & 0.53 & 0.58 & 0.62 & 0.53 & 0.54 \\
LlaMa-3.1-8B & 0.44 & 0.38 & 0.38 & 0.38 & 0.42 & 0.34 & 0.41 \\
starcoder2-7b & 0.51 & 0.47 & 0.47 & 0.47 & 0.51 & 0.43 & 0.47 \\
codegemma-7b & 0.56 & 0.48 & 0.48 & 0.50 & 0.51 & 0.48 & 0.54 \\
gemma-2-2b & 0.48 & 0.39 & 0.47 & 0.47 & 0.46 & 0.37 & 0.44 \\
Mistral-7B-v0.3 & 0.56 & 0.46 & 0.51 & 0.52 & 0.56 & 0.44 & 0.50 \\
Gemma 2 27b & 0.64 & 0.52 & 0.50 & 0.53 & 0.56 & 0.51 & 0.53 \\
LlaMa 70B & 0.51 & 0.47 & 0.41 & 0.45 & 0.45 & 0.46 & 0.44 \\
\bottomrule
\end{tabular}}
\end{table*}
%

\begin{figure*}[h]
    \centering
        \begin{subfigure}[t]{0.48\textwidth}
        \centering
        \includegraphics[width=0.8\textwidth]{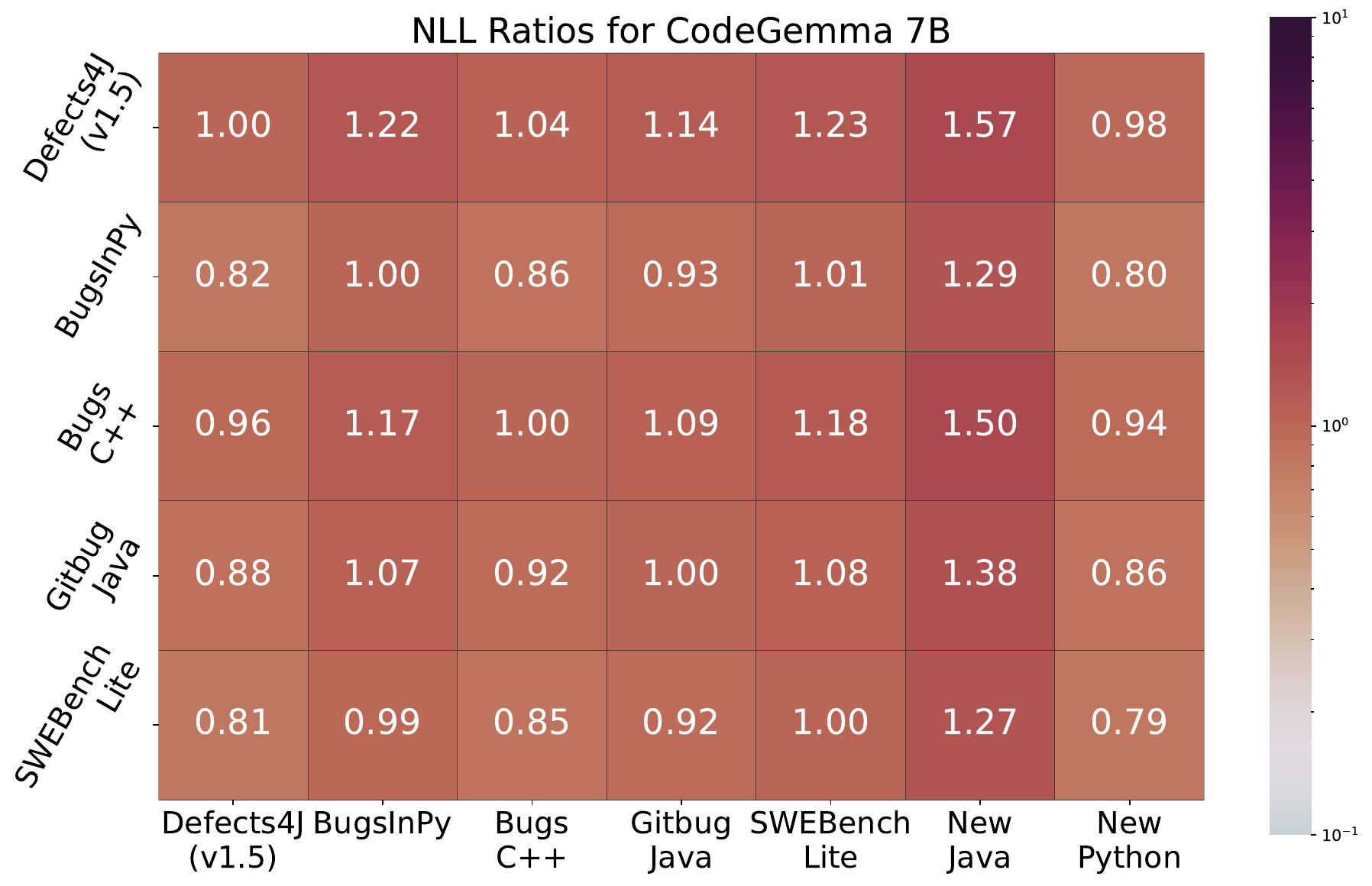}
        \caption{NLL ratios for CodeGemma 7B.}
    \end{subfigure}
    \hfill
    \begin{subfigure}[t]{0.48\textwidth}
        \centering
        \includegraphics[width=0.8\textwidth]
        {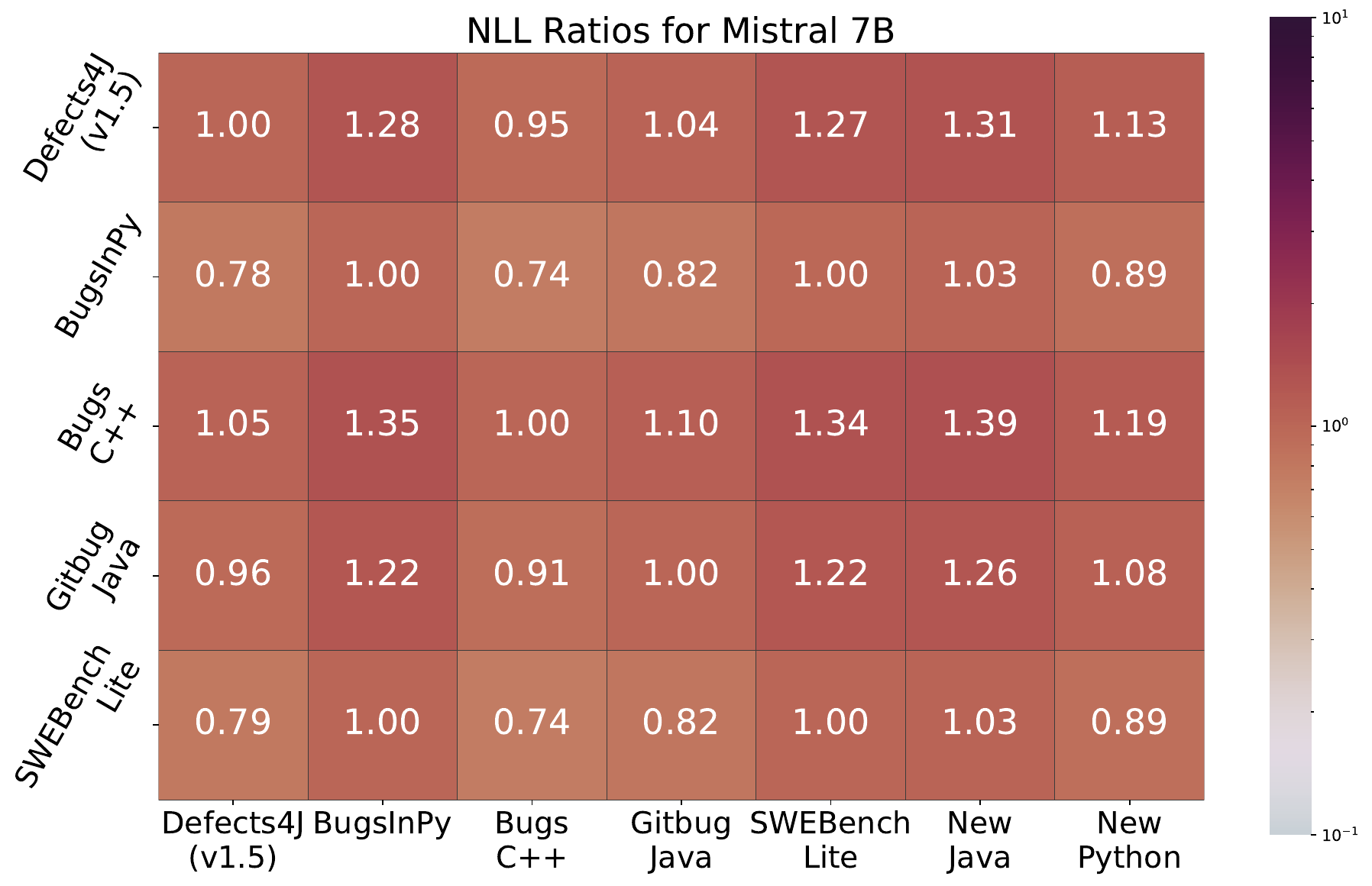}
        \caption{NLL ratios for Mistral 7B.    \vspace{1em}}
    \end{subfigure}

    \begin{subfigure}[t]{0.48\textwidth}
        \centering
        \includegraphics[width=0.8\textwidth]{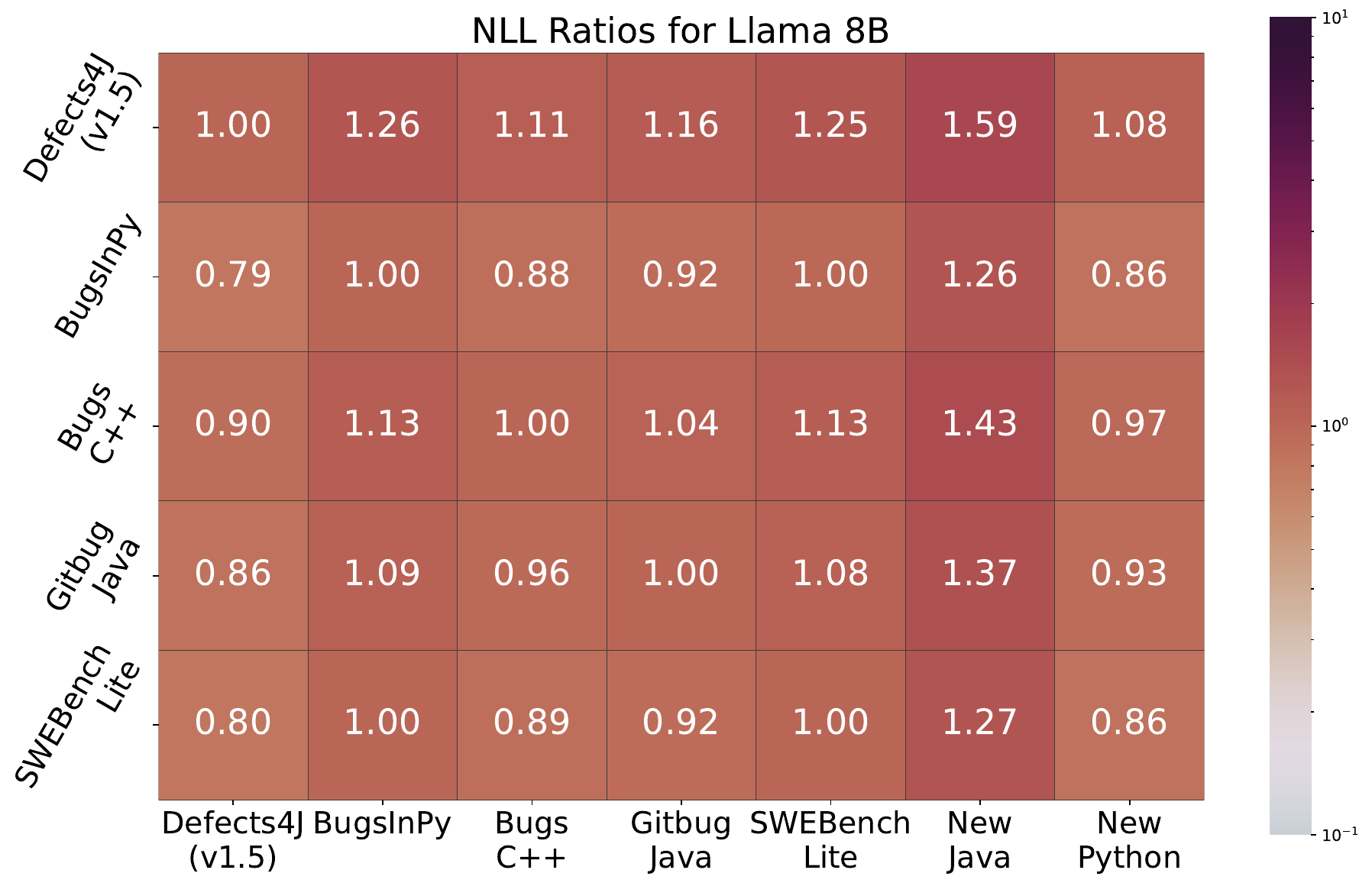}
        \caption{NLL ratios for LlaMa 3.1 8B.}
    \end{subfigure}
    \hfill
    \begin{subfigure}[t]{0.48\textwidth}
        \centering
g        \includegraphics[width=0.8\textwidth]{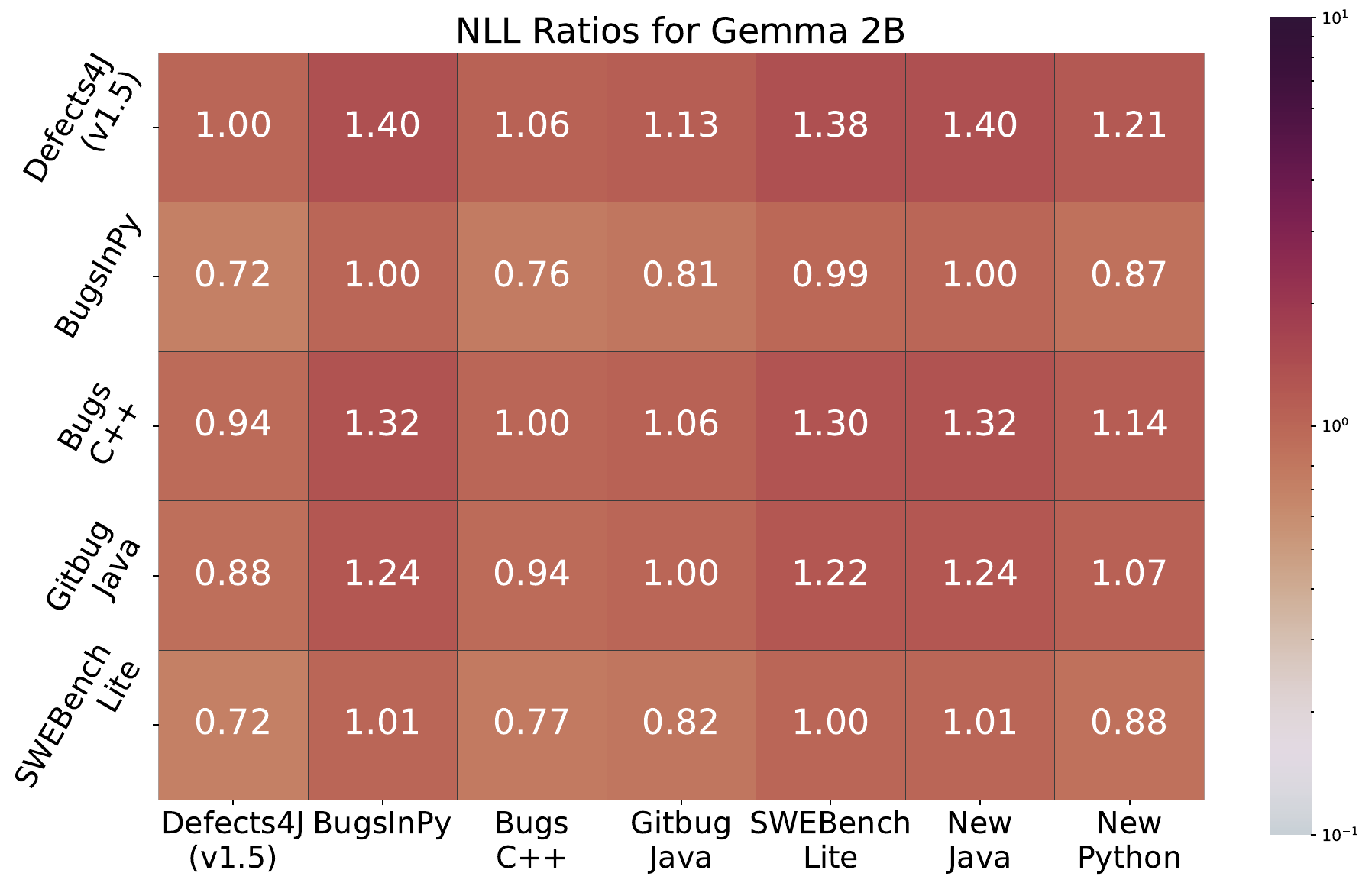}
        \caption{NLL ratios for Gemma 2B.}
    \end{subfigure}
    
    \vspace{1em} 
    \begin{subfigure}[t]{0.48\textwidth}
        \centering
        \includegraphics[width=0.8\textwidth]{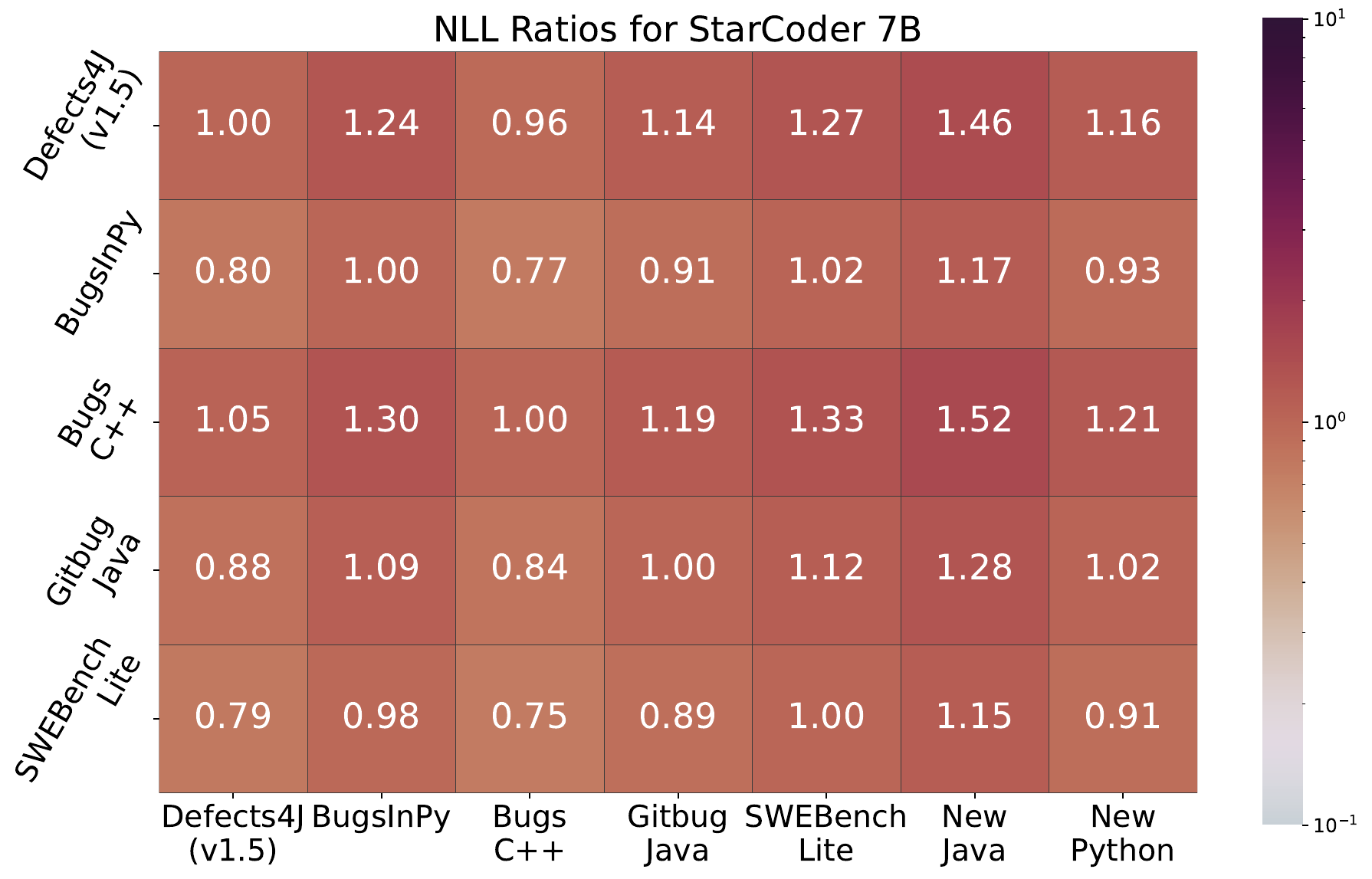}
        \caption{NLL ratios for StarCoder 7B.}
    \end{subfigure}
    \begin{subfigure}[t]{0.48\textwidth}
        \centering
        \includegraphics[width=0.8\textwidth]{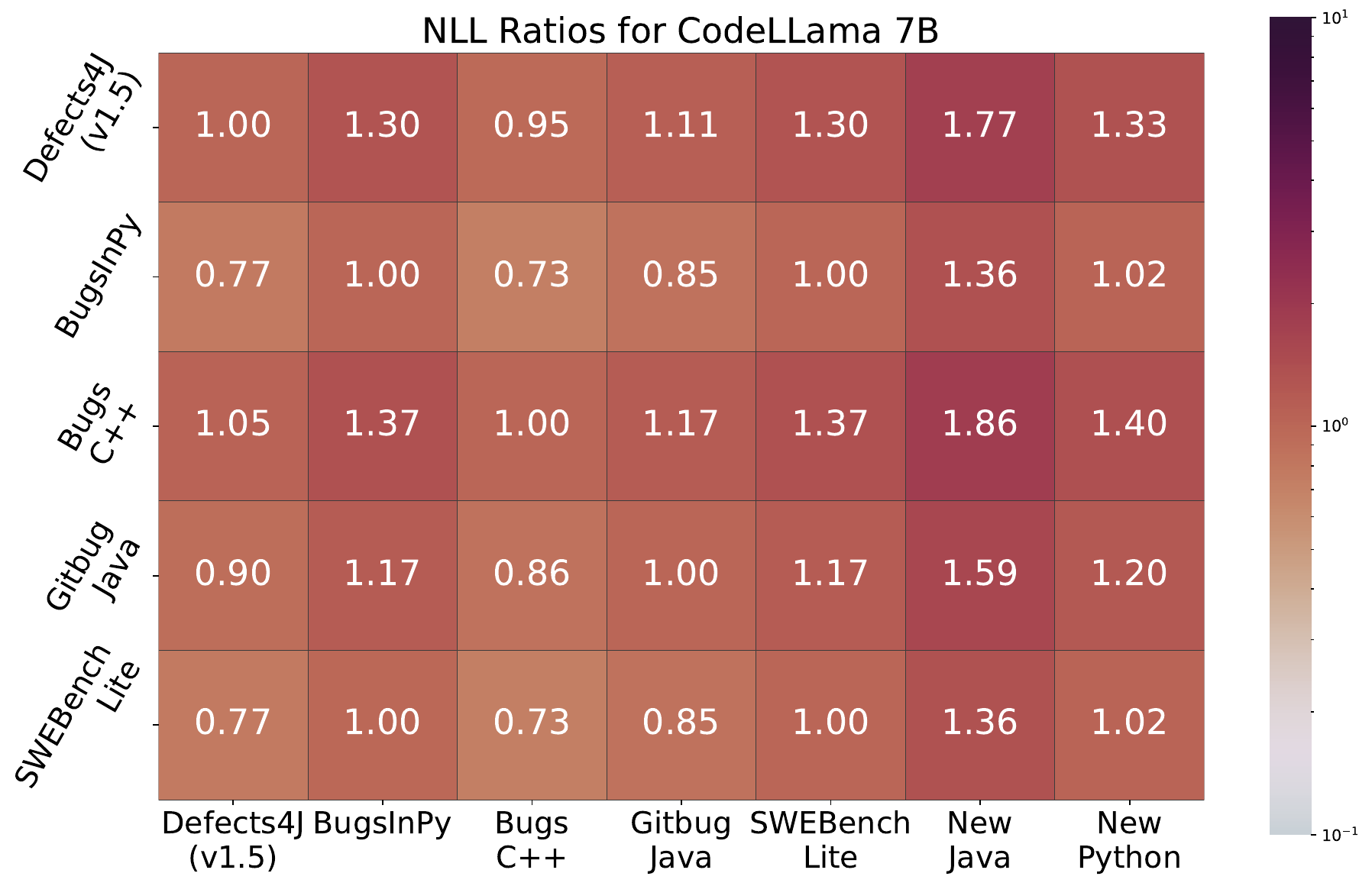}
        \caption{NLL ratios for CodeLLaMa 7B.}
    \end{subfigure}

    \caption{Heatmaps for \nll ratios}
    \label{fig:appendix-subfigures}
\end{figure*}

\begin{figure*}[h]
    \begin{subfigure}[t]{0.48\textwidth}
        \centering
        \includegraphics[width=0.9\textwidth]{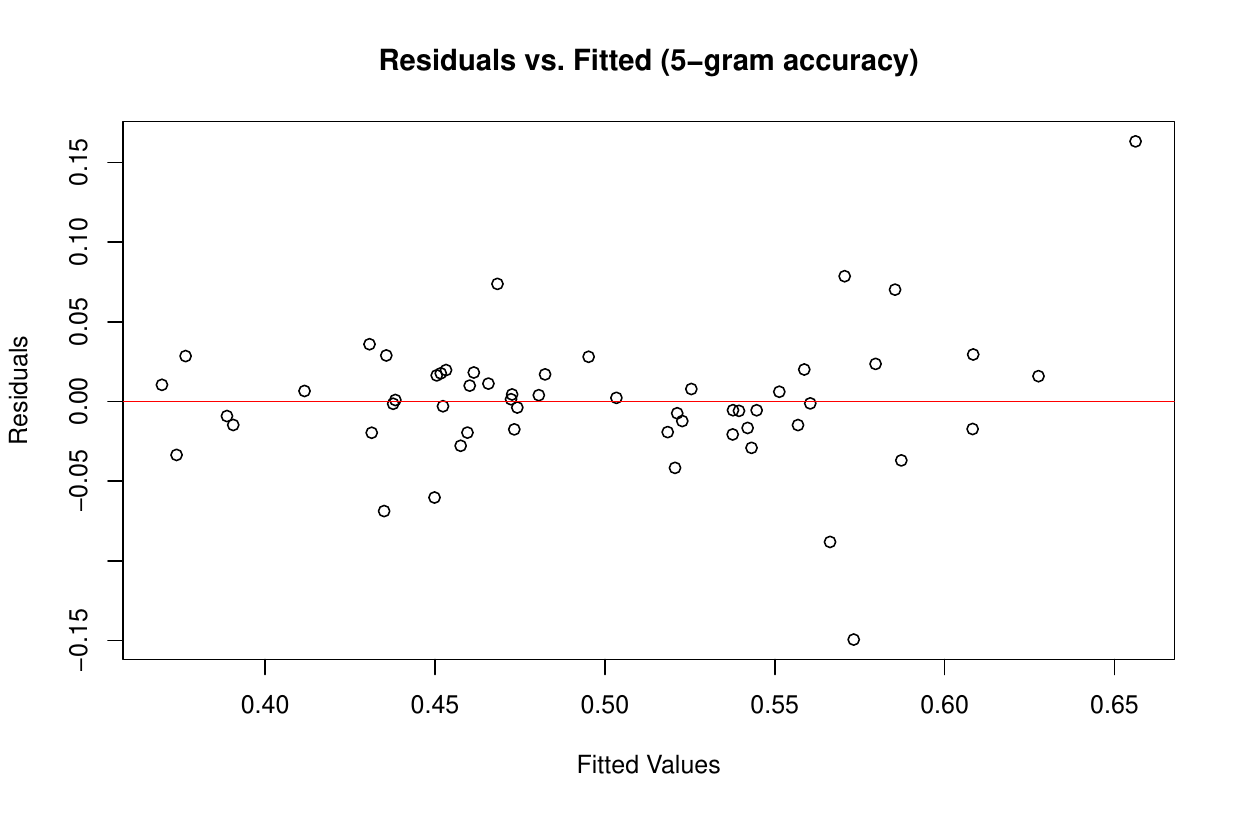}
        \caption{Residual vs. Fitted Plot for average \fivegram.}
    \end{subfigure}
    \hfill
    \begin{subfigure}[t]{0.48\textwidth}
        \centering
        \includegraphics[width=0.9\textwidth]{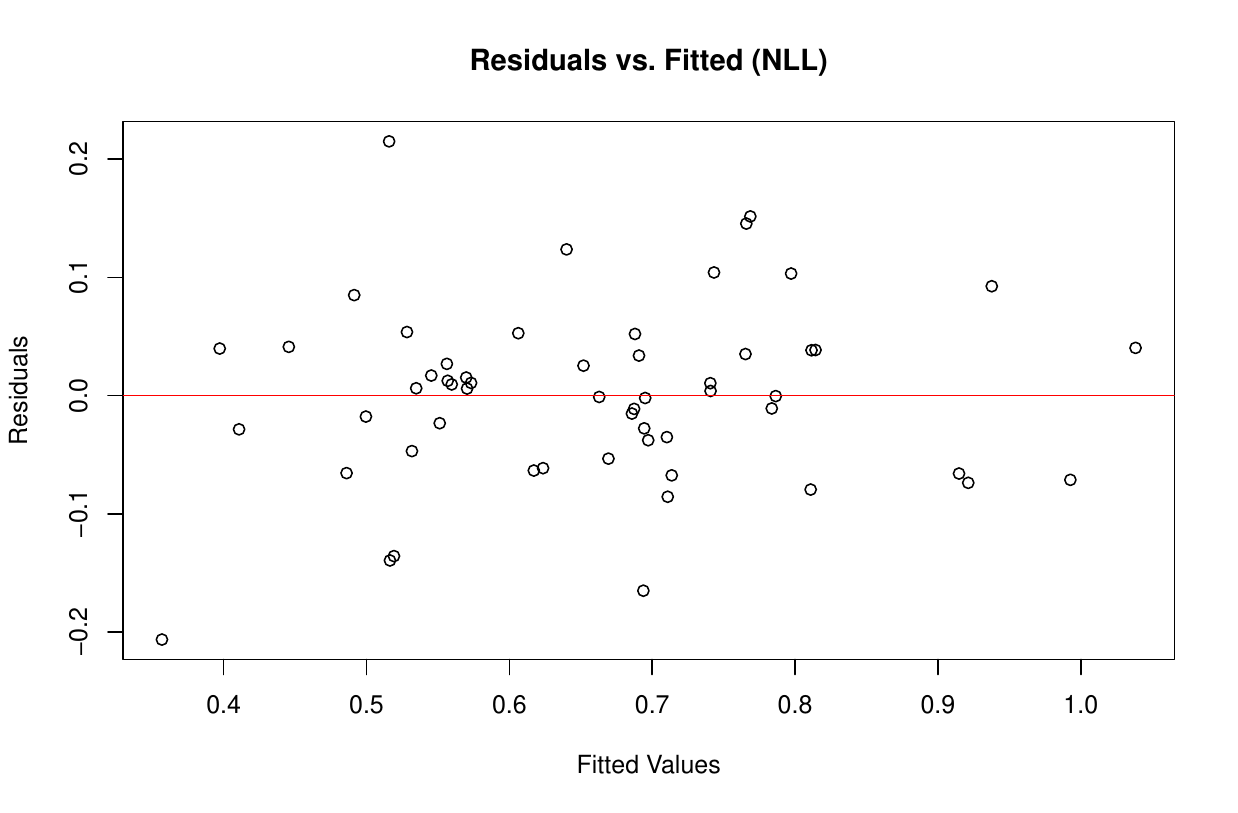}
        \caption{Residual vs. Fitted Plot for average \nll.}
    \end{subfigure}

        \caption{Residuals for the linear regressions.}
    \label{fig:residuals}
\end{figure*}

%% file: apr-data-leakage.bbl
\begin{thebibliography}{10}
\providecommand{\url}[1]{#1}
\csname url@rmstyle\endcsname
\providecommand{\newblock}{\relax}
\providecommand{\bibinfo}[2]{#2}
\providecommand\BIBentrySTDinterwordspacing{\spaceskip=0pt\relax}
\providecommand\BIBentryALTinterwordstretchfactor{4}
\providecommand\BIBentryALTinterwordspacing{\spaceskip=\fontdimen2\font plus
\BIBentryALTinterwordstretchfactor\fontdimen3\font minus \fontdimen4\font\relax}
\providecommand\BIBforeignlanguage[2]{{%
\expandafter\ifx\csname l@#1\endcsname\relax
\typeout{** WARNING: IEEEtran.bst: No hyphenation pattern has been}%
\typeout{** loaded for the language `#1'. Using the pattern for}%
\typeout{** the default language instead.}%
\else
\language=\csname l@#1\endcsname
\fi
#2}}

\bibitem{chen2021humaneval}
\BIBentryALTinterwordspacing
M.~Chen \emph{et~al.}, ``Evaluating large language models trained on code,'' \emph{CoRR}, vol. abs/2107.03374, 2021. [Online]. Available: \url{https://arxiv.org/abs/2107.03374}
\BIBentrySTDinterwordspacing

\bibitem{flSurvey}
W.~E. Wong \emph{et~al.}, ``A survey on software fault localization,'' \emph{IEEE Transactions on Software Engineering}, vol.~42, no.~8, pp. 707--740, 2016.

\bibitem{aprCACM}
C.~Le~Goues, M.~Pradel, and A.~Roychoudhury, ``Automated {P}rogram {R}epair,'' \emph{Communications of the ACM}, vol.~62, no.~12, pp. 56--65, 2019.

\bibitem{defects4j}
R.~Just, D.~Jalali, and M.~D. Ernst, ``Defects4j: a database of existing faults to enable controlled testing studies for java programs,'' in \emph{International Symposium on Software Testing and Analysis}, 2014, pp. 437--440.

\bibitem{bugsinpy}
R.~Widyasari \emph{et~al.}, ``Bugsinpy: a database of existing bugs in python programs to enable controlled testing and debugging studies,'' in \emph{Foundations of Software Engineering}, 2020.

\bibitem{swebench}
\BIBentryALTinterwordspacing
C.~E. Jimenez \emph{et~al.}, ``{SWE}-bench: Can language models resolve real-world github issues?'' in \emph{International Conference on Learning Representations}, 2024. [Online]. Available: \url{https://openreview.net/forum?id=VTF8yNQM66}
\BIBentrySTDinterwordspacing

\bibitem{cao2024concerneddatacontaminationassessing}
\BIBentryALTinterwordspacing
J.~Cao, W.~Zhang, and S.~Cheung, ``Concerned with data contamination? assessing countermeasures in code language model,'' \emph{CoRR}, vol. abs/2403.16898, 2024. [Online]. Available: \url{https://doi.org/10.48550/arXiv.2403.16898}
\BIBentrySTDinterwordspacing

\bibitem{kapoor2022leakagereproducibilitycrisismlbased}
\BIBentryALTinterwordspacing
S.~Kapoor and A.~Narayanan, ``Leakage and the reproducibility crisis in ml-based science,'' \emph{CoRR}, vol. abs/2207.07048, 2022. [Online]. Available: \url{https://doi.org/10.48550/arXiv.2207.07048}
\BIBentrySTDinterwordspacing

\bibitem{xu2024benchmarkingbenchmarkleakagelarge}
\BIBentryALTinterwordspacing
R.~Xu \emph{et~al.}, ``Benchmarking benchmark leakage in large language models,'' \emph{CoRR}, vol. abs/2404.18824, 2024. [Online]. Available: \url{https://doi.org/10.48550/arXiv.2404.18824}
\BIBentrySTDinterwordspacing

\bibitem{li2023estimatingcontaminationperplexityquantifying}
\BIBentryALTinterwordspacing
Y.~Li, ``Estimating contamination via perplexity: Quantifying memorisation in language model evaluation,'' \emph{CoRR}, vol. abs/2309.10677, 2023. [Online]. Available: \url{https://doi.org/10.48550/arXiv.2309.10677}
\BIBentrySTDinterwordspacing

\bibitem{bugscpp}
\BIBentryALTinterwordspacing
G.~An \emph{et~al.}, ``Bugsc++: A highly usable real world defect benchmark for c/c++,'' in \emph{International Conference on Automated Software Engineering}.\hskip 1em plus 0.5em minus 0.4em\relax IEEE Press, 2024, p. 2034–2037. [Online]. Available: \url{https://doi.org/10.1109/ASE56229.2023.00208}
\BIBentrySTDinterwordspacing

\bibitem{gitbug-java}
A.~Silva, N.~Saavedra, and M.~Monperrus, ``Gitbug-java: A reproducible benchmark of recent java bugs,'' in \emph{International Conference on Mining Software Repositories}, ser. MSR ’24.\hskip 1em plus 0.5em minus 0.4em\relax ACM, 2024, p. 118–122.

\bibitem{xia2024agentlessdemystifyingllmbasedsoftware}
\BIBentryALTinterwordspacing
C.~S. Xia \emph{et~al.}, ``Agentless: Demystifying llm-based software engineering agents,'' \emph{CoRR}, vol. abs/2407.01489, 2024. [Online]. Available: \url{https://doi.org/10.48550/arXiv.2407.01489}
\BIBentrySTDinterwordspacing

\bibitem{wang2024openhandsopenplatformai}
\BIBentryALTinterwordspacing
X.~Wang \emph{et~al.}, ``Openhands: An open platform for ai software developers as generalist agents,'' 2024. [Online]. Available: \url{https://arxiv.org/abs/2407.16741}
\BIBentrySTDinterwordspacing

\bibitem{aleithan2024swebenchenhancedcodingbenchmark}
\BIBentryALTinterwordspacing
R.~Aleithan \emph{et~al.}, ``Swe-bench+: Enhanced coding benchmark for llms,'' 2024. [Online]. Available: \url{https://arxiv.org/abs/2410.06992}
\BIBentrySTDinterwordspacing

\bibitem{broderMinHash}
A.~Broder, ``On the resemblance and containment of documents,'' in \emph{Compression and Complexity of SEQUENCES 1997 (Cat. No.97TB100171)}, 1997, pp. 21--29.

\bibitem{gionisLSH}
A.~Gionis, P.~Indyk, and R.~Motwani, ``Similarity search in high dimensions via hashing,'' in \emph{International Conference on Very Large Data Bases}, ser. VLDB '99, 1999, p. 518–529.

\bibitem{nijkamp2022codegen}
E.~Nijkamp \emph{et~al.}, ``Codegen: An open large language model for code with multi-turn program synthesis,'' in \emph{International Conference on Learning Representations}.\hskip 1em plus 0.5em minus 0.4em\relax OpenReview.net, 2023.

\bibitem{rozière2024codellamaopenfoundation}
\BIBentryALTinterwordspacing
B.~Rozière \emph{et~al.}, ``Code llama: Open foundation models for code,'' 2023. [Online]. Available: \url{https://doi.org/10.48550/arXiv.2308.12950}
\BIBentrySTDinterwordspacing

\bibitem{touvron2023llama}
\BIBentryALTinterwordspacing
H.~Touvron \emph{et~al.}, ``Llama: Open and efficient foundation language models,'' 2023. [Online]. Available: \url{https://doi.org/10.48550/arXiv.2302.13971}
\BIBentrySTDinterwordspacing

\bibitem{lozhkov2024starcoder2stackv2}
A.~Lozhkov \emph{et~al.}, ``Starcoder 2 and the stack v2: The next generation,'' 2024.

\bibitem{gemmateam2024gemma2}
\BIBentryALTinterwordspacing
G.~Team \emph{et~al.}, ``Gemma 2: Improving open language models at a practical size,'' 2024. [Online]. Available: \url{https://doi.org/10.48550/arXiv.2408.00118}
\BIBentrySTDinterwordspacing

\bibitem{codegemmateam2024codegemma}
\BIBentryALTinterwordspacing
C.~Team \emph{et~al.}, ``Codegemma: Open code models based on gemma,'' 2024. [Online]. Available: \url{https://doi.org/10.48550/arXiv.2406.11409}
\BIBentrySTDinterwordspacing

\bibitem{jiang2023mistral7b}
\BIBentryALTinterwordspacing
A.~Q. Jiang \emph{et~al.}, ``Mistral 7b,'' 2023. [Online]. Available: \url{https://arxiv.org/abs/2310.06825}
\BIBentrySTDinterwordspacing

\bibitem{yang2023largelanguagemodelstestfree}
A.~Z.~H. Yang, C.~{Le Goues}, R.~Martins, and V.~J. Hellendoorn, ``Large language models for test-free fault localization,'' in \emph{International Conference on Software Engineering}.\hskip 1em plus 0.5em minus 0.4em\relax {ACM}, 2024, pp. 17:1--17:12.

\bibitem{silva2024repairllamaefficientrepresentationsfinetuned}
\BIBentryALTinterwordspacing
A.~Silva, S.~Fang, and M.~Monperrus, ``Repairllama: Efficient representations and fine-tuned adapters for program repair,'' \emph{CoRR}, 2023. [Online]. Available: \url{https://doi.org/10.48550/arXiv.2312.15698}
\BIBentrySTDinterwordspacing

\bibitem{zhou2024largelanguagemodelvulnerability}
X.~Zhou, T.~Zhang, and D.~Lo, ``Large language model for vulnerability detection: Emerging results and future directions,'' in \emph{International Conference on Software Engineering: New Ideas and Emerging Results}.\hskip 1em plus 0.5em minus 0.4em\relax {ACM}, 2024, pp. 47--51.

\bibitem{kocetkov2022stack3tbpermissively}
\BIBentryALTinterwordspacing
D.~Kocetkov \emph{et~al.}, ``The stack: 3 {TB} of permissively licensed source code,'' \emph{Trans. Mach. Learn. Res.}, 2023. [Online]. Available: \url{https://openreview.net/forum?id=pxpbTdUEpD}
\BIBentrySTDinterwordspacing

\bibitem{gabel2010Uniqueness}
M.~Gabel and Z.~Su, ``A study of the uniqueness of source code,'' in \emph{International Symposium on Foundations of Software Engineering}, G.~Roman and A.~van~der Hoek, Eds.\hskip 1em plus 0.5em minus 0.4em\relax {ACM}, 2010, pp. 147--156.

\bibitem{luo2024empiricalstudycatastrophicforgetting}
\BIBentryALTinterwordspacing
Y.~Luo \emph{et~al.}, ``An empirical study of catastrophic forgetting in large language models during continual fine-tuning,'' \emph{CoRR}, vol. abs/2308.08747, 2023. [Online]. Available: \url{https://doi.org/10.48550/arXiv.2308.08747}
\BIBentrySTDinterwordspacing

\bibitem{tirumala2022memorizationoverfittinganalyzingtraining}
K.~Tirumala, A.~H. Markosyan, L.~Zettlemoyer, and A.~Aghajanyan, ``Memorization without overfitting: Analyzing the training dynamics of large language models,'' in \emph{Advances in Neural Information Processing Systems 35: Annual Conference on Neural Information Processing Systems}, 2022.

\bibitem{rajpurkar2016squad100000questionsmachine}
\BIBentryALTinterwordspacing
P.~Rajpurkar, J.~Zhang, K.~Lopyrev, and P.~Liang, ``Squad: 100, 000+ questions for machine comprehension of text,'' in \emph{Conference on Empirical Methods in Natural Language Processing}, 2016, pp. 2383--2392. [Online]. Available: \url{https://doi.org/10.18653/v1/d16-1264}
\BIBentrySTDinterwordspacing

\bibitem{hendrycks2021measuringmassivemultitasklanguage}
\BIBentryALTinterwordspacing
D.~Hendrycks \emph{et~al.}, ``Measuring massive multitask language understanding,'' in \emph{International Conference on Learning Representations1}.\hskip 1em plus 0.5em minus 0.4em\relax OpenReview.net, 2021. [Online]. Available: \url{https://openreview.net/forum?id=d7KBjmI3GmQ}
\BIBentrySTDinterwordspacing

\bibitem{chen2021evaluatinglargelanguagemodels}
\BIBentryALTinterwordspacing
M.~Chen \emph{et~al.}, ``Evaluating large language models trained on code,'' 2021. [Online]. Available: \url{https://arxiv.org/abs/2107.03374}
\BIBentrySTDinterwordspacing

\bibitem{austin2021programsynthesislargelanguage}
\BIBentryALTinterwordspacing
J.~Austin \emph{et~al.}, ``Program synthesis with large language models,'' \emph{CoRR}, 2021. [Online]. Available: \url{https://arxiv.org/abs/2108.07732}
\BIBentrySTDinterwordspacing

\bibitem{hendrycks2021measuringcodingchallengecompetence}
D.~Hendrycks \emph{et~al.}, ``Measuring coding challenge competence with {APPS},'' in \emph{Neural Information Processing Systems Track on Datasets and Benchmarks}, J.~Vanschoren and S.~Yeung, Eds., 2021.

\bibitem{xia2023conversationgoingfixing162}
\BIBentryALTinterwordspacing
C.~S. Xia and L.~Zhang, ``Keep the conversation going: Fixing 162 out of 337 bugs for {\textdollar}0.42 each using chatgpt,'' \emph{CoRR}, 2023. [Online]. Available: \url{https://doi.org/10.48550/arXiv.2304.00385}
\BIBentrySTDinterwordspacing

\bibitem{shi2024detectingpretrainingdatalarge}
\BIBentryALTinterwordspacing
W.~Shi \emph{et~al.}, ``Detecting pretraining data from large language models,'' in \emph{International Conference on Learning Representations}.\hskip 1em plus 0.5em minus 0.4em\relax OpenReview.net, 2024. [Online]. Available: \url{https://openreview.net/forum?id=zWqr3MQuNs}
\BIBentrySTDinterwordspacing

\bibitem{jain2024livecodebenchholisticcontaminationfree}
\BIBentryALTinterwordspacing
N.~Jain \emph{et~al.}, ``Livecodebench: Holistic and contamination free evaluation of large language models for code,'' \emph{CoRR}, vol. abs/2403.07974, 2024. [Online]. Available: \url{https://doi.org/10.48550/arXiv.2403.07974}
\BIBentrySTDinterwordspacing

\bibitem{zuoLeakage}
C.~Zuo, Z.~Lin, and Y.~Zhang, ``Why does your data leak? uncovering the data leakage in cloud from mobile apps,'' in \emph{Symposium on Security and Privacy}, 2019, pp. 1296--1310.

\end{thebibliography}
